\documentclass{emulateapj}
\usepackage{graphicx}
\usepackage{amsfonts}
\usepackage{pifont}
\newcommand{\tickNo}{\hspace{1pt}\ding{55}}

\def\aap{\mbox{A\&A}}
\def\apjl{\mbox{ApJ}}
\def\apjs{\mbox{ApJS}}

\def\ms{\mbox{$M_{\rm *}$}}

\def\mh{\mbox{$M_{\rm h}$}}

\def\cmfsat{\mbox{$\Phi_{\rm sat}(\ms|\mh)$}}

\def\cmf{\mbox{CSMF}}
\def\gsmf{\mbox{GSMF}}
\def\hmf{\mbox{HMF}}
\def\phig{\mbox{$\phi_g$}}
\def\phih{\mbox{$\phi_{\rm h}$}}
\def\phicen{\mbox{$\phi_{g,{\rm  cen}}$}}

\def\NG{\mbox{$\langle N(>\ms|\mh)\rangle$}}
\def\Ns{\mbox{$\langle N_s(>\ms|\mh)\rangle$}}
\def\Nsat{\mbox{$\langle N_s\rangle$}}
\def\Nc{\mbox{$\langle N_c(>\ms|\mh)\rangle$}}

\def\ng{\mbox{$n_g$}}
\def\ngcen{\mbox{$n_{g,\rm cen}$}}
\def\ngsat{\mbox{$n_{g,\rm sat}$}}
\def\mc{\mbox{$M_{\rm *,c}(\mh)$}}
\def\mcs{\mbox{$M_{\rm *,s}(\msub)$}}

\def\xigg{\mbox{$\xi_{\rm gg}(r)$}}
\def\xiggh{\mbox{$\xi_{\rm gg}^{\rm 1h}(r)$}}
\def\xigghh{\mbox{$\xi_{\rm gg}^{\rm 2h}(r)$}}
\def\wp{\mbox{$w_{\rm p}(r_{\rm p})$}}
\def\w{\mbox{$w_{\rm p}$}}

\def\fs{\mbox{$f_{\rm *}$}}

\def\msun{\mbox{M$_\odot$}}

\def\avec{\mbox{$\mathbf{a}$}}

\def\Mstar{\mbox{$\mathcal{M}^*$}}

\def\amt{\mbox{AMT}}
\def\shmr{\mbox{CHMR}}
\def\ssmr{\mbox{SSMR}}

\def\pcf{\mbox{2PCF}}
\def\scmf{\mbox{SubhCMF}}

\def\msub{\mbox{$m_{\rm sub}$}}
\def\macc{\mbox{$m_{\rm acc}$}}
\def\mobs{\mbox{$m_{\rm obs}$}}

\def\lcdm{\mbox{$\Lambda$CDM}}

\def\Nsub{\mbox{$\langle N_{\rm sub}(>\msub|\mh)\rangle$}}

\def\phisub{\mbox{$\Phi_{\rm sub}(\msub|\mh)$}}
\def\Psub{\mbox{$P_{\rm sat}(\ms|\msub)$}}
\def\Pcen{\mbox{$P_{\rm cen}(\ms|\mh)$}}

\def\phisat{\mbox{$\phi_{g,{\rm sat}}$}}
\def\densub{\mbox{$\phi_{\rm sub}$}}

\defcitealias{YMB09}{YMB09}
\defcitealias{Yang2011}{Y12} 
\defcitealias{Baldry08}{BGD08}
\defcitealias{LW09}{LW09}
\defcitealias{LW11}{LW11}
\defcitealias{Boylan-Kolchin+10}{BK+10}
\defcitealias{RDA12}{RDA12}
\def\ltsima{$\; \buildrel < \over \sim \;$}    
\def\lesssim{\lower.5ex\hbox{\ltsima}}           
\def\gtsima{$\; \buildrel > \over \sim \;$}    
\def\grtsim{\lower.5ex\hbox{\gtsima}}           

\slugcomment{Submitted to ApJ}

\shorttitle{Galaxy--halo/subhalo connection}
\shortauthors{Rodriguez-Puebla, Avila-Reese \& Drory}

\begin{document}


\title{The galaxy--halo/subhalo connection: mass relations and implications
 for some satellite occupational distributions}


\author{A. Rodr\'iguez-Puebla, V. Avila-Reese and N. Drory}
\affil{Instituto de Astronom\'ia, 
Universidad Nacional Aut\'onoma de M\'exico,
A. P. 70-264, 04510, M\'exico, D.F., M\'exico.}

\email{apuebla@astro.unam.mx}



\begin{abstract} 
  We infer the local stellar-to-halo/subhalo mass relations (MRs) for
  central and satellite galaxies \textit{separately}.  Our statistical
  method is extending the abundance matching, halo occupation
  distribution, and conditional stellar mass function formalisms. We
  constrain the model using several combinations of observational
  data, consisting of the total galaxy stellar mass function (\gsmf),
  its decomposition into centrals and satellites, and the projected
  two-point correlation functions (\pcf s) measured in different
  stellar mass (\ms) bins. In addition, we use the \lcdm\ halo and
  subhalo mass functions. The differences among the resulting MRs are
  within the model-fit uncertainties (which are very small, smaller
  than the intrinsic scatter between galaxy and halo mass), no matter
  what combination of data are used. This shows that matching
  abundances or occupational numbers is equivalent, and that the \gsmf
  s and \pcf s are tightly connected. We also constrain the values of
  the intrinsic scatter around the central-halo (CH) and
  satellite-subhalo (SS) MRs assuming them to be constant: $\sigma_c =
  0.168\pm 0.051$ dex and $\sigma_s = 0.172\pm0.057$ dex,
  respectively.  The CH and SS MRs are actually different, in
  particular when we take the subhalo mass at the present-day epoch
  instead of at their accretion time. When using the MRs for studying
  the satellite population (e.g., in the Milky Way, MW), the SS MR
  should be chosen instead of the average one. Our model allows one to
  calculate several population statistics. We find that the central
  galaxy \ms\ is not on average within the mass distribution of the
  most-massive satellite, even for cluster-sized halos, i.e., centrals
  are not a mere realization of the high-end of the satellite mass
  function; however for $> 3\times 10^{13}$ \msun\ halos, $\sim 15\%$
  of centrals could be. We also find that the probabilities of
  MW-sized halos of having $N$ Magellanic-Clouds (MCs)-sized
  satellites agree well with observational measures; for a halo mass
  of $2\times 10^{12}$ \msun, the probability to have 2 MCs is
  $5.4\%$, but if we exclude those systems with satellites larger than
  the MCs, then the probability decreases to $<2.2\%$.
\end{abstract}


\section{Introduction}
The statistical description of the galaxy population is a valuable tool
for understanding the properties of galaxies and the way they cluster,
as well as the role that mass and environment play in shaping
these properties.  Moreover, statistical descriptors such as the luminosity
function, the galaxy stellar mass function (\gsmf), and the
two-point correlation function (\pcf) has allowed us to probe galaxy
evolution and its connection to the cosmological initial conditions of
structure formation \citep[e.g.,][]{Peebles1980,Yoo+2009}.
Such a connection is of vital importance in studies devoted to
the development of the current \lcdm\ cosmological paradigm.  A key
ingredient in these studies is the link between galaxy and dark
matter halo properties.  Such a link allows to project the theoretical
dark matter halo population onto the observable galaxies.

Recently, progress towards connecting galaxies and halos has been made
through the development of several techniques for observationally
estimating the dark halo masses of luminous galaxies, such as weak
lensing \citep{Mandelbaum06,Mandelbaum08,Schulz10}, kinematics of
satellite galaxies \citep{Conroy+2007,More09,More11,Wojtak&Mamon12},
and galaxy clusters \citep{Yang+07,YMB09,Hansen09,Yang2011}.  However,
these direct probes of halo mass still have large uncertainties.

Consequentially, semi-empirical approaches that link the galaxy and
dark matter halo distributions statistically are of great importance.
For example, the Halo Occupation Distribution (HOD) formalism, which
describes the probability for finding $N$ galaxies in halos of mass
\mh, has been used successfully to understand the non-linear relation
between the distribution of galaxies and matter, for instance, at the
level of the power spectra
\citep{Seljak2000,Peacock-Smith00,Cooray-Sheth03,Yoo+2009}, or the
two-point correlation functions \citep[][and references
therein]{Berlind-Weinberg02,
  Cooray-Sheth03,Zehavi05,Abbas+2006,Tinker+2008,Zehavi11,Watson+2011,Watson+2012,Wake+2012}.

However, the HOD model provides only information on the total number
of galaxies above some luminosity or stellar mass threshold per halo,
and constrains only the halo mass of the central galaxy.  In order to
describe the detailed halo occupation and mass distribution of central
and satellite galaxies, \citet{Yang+03} introduced the conditional
luminosity (or stellar mass) function (CSMF) in the HOD model (see
also e.g., \citealp{YMB09}, hereafter \citetalias{YMB09},
\citealp{Moster10, Leauthaud11A, Leauthaud11B,Yang2011}).  The CSMF is
defined as the average number of galaxies with stellar masses between
$\ms\pm d\ms/2$ occupying a halo of a given mass \mh.  Nevertheless,
both the HOD model and the CSMF formalism assume a parametric
description for the satellite population distributions which is
constrained using observations.

In order to avoid an arbitrary parametric description for the
satellite population, the above models can be generalized with the
abundance matching technique (hereafter, \amt; \citealp[e.g.,]
[]{Vale+04,Kravtsov04,Conroy+06,Shankar+06,Weinberg+2008,Baldry08,
  ConroyWechsler09,Drory+09,Moster10,Behroozi10,Guo10,
  Behroozi+12,Reddick+2012,Papastergis+12}).  Under the hypothesis
that there exists a one-to-one monotonic relation between stellar mass
and (sub)halo mass, the matching the total galaxy and halo
plus subhalo abundances yields a {\it global} (average) relation
between \ms\ and \mh. Note that in this simple procedure, the
central-to-halo and satellite-to-subhalo mass relations (hereafter
\shmr\ and \ssmr, respectively) are not
differentiated. 
Recently, \citet{Simha+2012} have found in their cosmological
N-body/hydrodynamics 
simulations that both mass relations are nearly identical if the
subhalo masses, \msub, are defined at their accretion times.
Additionally, previous studies have shown that when the \amt\ results
are applied to the HOD model with \msub\ defined at the accretion
epoch, then the spatial clustering of galaxies is mostly recovered
(e.g.,\citealp{Conroy+06,Moster10}).  Similar results are expected
when \msub\ is defined at the observation time but a global offset is
applied to account for the average effect of subhalo mass loss due to
tidal stripping \citep{Vale+04,Weinberg+2008}.

On the other hand, there is no reason to assume a priori the \ssmr\ to
be identical to the \shmr\ (\citealp{Neistein11}, \citealp{RDA12},
hereafter \citetalias{RDA12}).  For accretion-time \msub, such
an assumption implies that the change of the stellar masses of
satellites after their accretion will be such that they would occupy
the $z=0$ central-to-halo mass relation or, more generally, that the
\shmr\ almost does not change with time. 
Recent studies based on large halo-based group catalogs
\citep[e.g.,][]{Wetzel+12} or on the predicted bulge-to-total mass
ratio of central galaxies\citep{Zavala+12} have shown that once
satellite galaxies are accreted, they evolve roughly as a central
galaxy at least for several Gyrs. This could imply that the \ssmr\
with \msub\ defined at accretion time may not be equal to the $z=0$
\shmr. Nevertheless, in the cosmological simulations of
\citet{Simha+2012}, despite the fact that satellites continue to grow
after accretion, both mass relations end up similar.  It is therefore
likely that the growth in mass as well as the change of the \shmr\
with time are very small.

In \citetalias{RDA12} we extended the \amt\ to determine the \shmr\ and \ssmr\ separately, 
using the observed decomposition of the \gsmf s into centrals and satellites. We
have found that indeed the \ssmr\ is not equal to the \shmr, and that applying them
to the HOD + \cmf\ model leads to satellite \cmf\ and correlation functions 
in excellent agreement with observational data. 
Actually, when \msub\ is defined at the accretion time, the $z=0$ mass
relations become close but not equal (see Fig. 2 in
\citetalias{RDA12}).  Additionally, \citetalias{RDA12} show that the
uncertainty in the \amt\ related to the satellite stellar mass growth
can be avoided if subhalo masses are defined at the time of
observation rather than at the time of accretion.  \citetalias{RDA12}
also suggest that the central-halo and satellite-subhalo mass
relations can be determined simultaneously using the correlation
functions as observational input, instead of the \gsmf\ decomposed
into satellites and centrals. This is presumably because {\it matching
  abundances of satellite to subhalos is essentially equivalent to
  matching their corresponding occupational numbers (and vice versa)}.

In the present paper, we aim to test the above statements.  We will
also probe how robust the determinations of the central-to-halo and
satellite-to-subhalo mass relations through our extended \amt\ and
HOD+\cmf\ combined model are. We will explore whether these mass
relations vary significantly depending on the combinations of
observational data being used; in particular, we will explore whether
the uncertainties in the model parameters that describe the mass
relations shrink significantly when more observational constraints are
added.

Our model proves to be a powerful tool for connecting the \lcdm\
(sub)halo statistics to the statistics of the central/satellite galaxy
populations. In this sense, one may predict many halo occupational
distributions and probabilities as a function of scale; for instance,
the mass distribution of the most massive satellites as a function of
halo mass or the probability of a halo hosting $N$ satellites in a
given stellar mass range or, more generally, the whole satellite
\cmf. We will discuss some results obtained for these
occupational distributions and probabilities.

In Section 2 we describe our extended \amt\ and HOD+\cmf\ model, and
present the different combinations of data to be used to constrain the
model parameters.  The results of our model for the different data
sets are presented in Section 3. In particular, we compare the
central-halo and satellite-subhalo mass relations obtained using
different data sets. We also constrain the intrinsic scatter around
the mean central-halo and satellite-subhalo mass relations.  In \S 4
we discuss the halo occupational statistics related to the halo mass
dependence of the satellite \cmf, the stellar mass gap between the
most massive satellite and the central galaxy, and the probabilities
of Milky-Way (MW) sized halos having 1, 2, or more Magellanic
Cloud-sized satellites. Section 5 is devoted to discuss the robustness
of the obtained mass relations and their model uncertainties, as well
as the implications of extrapolating our obtained \ssmr\ to masses as
small as the MW dwarf spheroidal galaxies.  Finally, we present our
conclusions in Section 6.
 
We adopt cosmological parameter values close to WMAP 7:
$\Omega_\Lambda=0.73,\Omega_{\rm M}=0.27, h=0.70, n_s=0.98$ 
and $\sigma_8=0.84$.

\section{Methodology}

In the following we present our model connecting galaxies to halos and
subhalos via their occupational numbers. This is done under the
assumption that on average the central-to-halo and
satellite-to-subhalo relations are monotonic.  The model relates in a
self-consistent way the \gsmf\ decomposed into centrals and
satellites, the \lcdm\ halo/subhalo mass functions, the satellite \cmf
s, and the galaxy projected \pcf s.  As a result, it constrains both
the \shmr\ and the \ssmr, and predicts the satellite \cmf\ and several
other occupational statistics.  Unlike previous models of this kind
\citep[e.g.,][]{Moster10}, the \shmr\ and \ssmr\ are treated
separately.

\subsection{Connecting galaxies to halos and subhalos}

The \emph{total} \gsmf\ is decomposed into satellites and central
galaxies,
\begin{equation}
  \phig(\ms)=\phicen(\ms)+\phisat(\ms),
\end{equation}
which after integration yields the mean cumulative number density of
galaxies with stellar masses greater than \ms,
\begin{equation}
  \int_{M_*}^{\infty}\phig d\ms'=\int_{M_*}^{\infty}\phicen d\ms'
  +\int_{M_*}^{\infty}\phisat d\ms',
\end{equation}
or, in short, 
\begin{equation}
  \ng(>\ms)=\ngcen(>\ms)+\ngsat(>\ms).
\end{equation}

\subsubsection{Central galaxies}

For constructing the central \gsmf, we will use the conditional
probability that a given halo of mass \mh\ is inhabited by a central
galaxy with stellar mass between $\ms\pm d\ms/2$, $\Pcen d\ms$, and
assume this distribution to be log-normal:
\begin{eqnarray}
  \Pcen d\ms=\frac{d\ms}{\sqrt{2\pi\sigma_c^2}\ms\ln(10)}\times& & \nonumber \\
  \exp\left[{-\frac{\log^2(\ms/\mc)}{2\sigma_c^2}}\right],
  \label{cmfc}
\end{eqnarray}
with $\sigma_c$ being the intrinsic scatter (width), expressed in dex
units, around $\log\mc$, 
the mean stellar-to-halo mass relation of \emph{central galaxies} (\shmr). Formally,
\Pcen\ maps the halo mass function, \hmf, onto the central \gsmf, thereby encoding all
the physical processes involved in galaxy formation inside the halos. We
parametrize $\log\mc$ using the functional form proposed by \citet{Behroozi+12},
\begin{eqnarray}
  \log\mc=\log(\epsilon_c M_{1,c})+f(\log(\mh/M_{1,c}))-f(0),
  \label{msmh}
\end{eqnarray}
where
\begin{equation}
  f(x)=\delta_c\frac{(\log(1+e^x))^{\gamma_c}}{1+e^{10^{-x}}}-\log(10^{\alpha_c x}+1).
\end{equation}
This function behaves as power law with slope $\alpha$ at masses much
smaller than $M_{1,c}$, and as a sub-power law with slope $\gamma_c$ at
larger masses.  This parametrization maps the \lcdm\ \hmf\ to a
Schechter-like \gsmf\ \citep{Schechter1976}.

The mean number density of central galaxies with stellar masses between $\ms\pm
d\ms/2$, (i.e., the central \gsmf) is given by
\begin{equation}
  \phicen(\ms)d\ms=d\ms\int_0^{\infty}\Pcen\phih(\mh) d\mh,
  \label{phig}
\end{equation}
where $\phih(\mh)$ is the \emph{distinct} HMF. We use the fitted results to the {\it distinct} HMF 
from cosmological simulations carried out in \citet{Tinker+08} as reported in their Appendix B. Here we
define halo masses at the virial radius, i.e.\ the halo radius where 
the spherical overdensity is $\Delta_{\rm vir}$ 
times the mean matter density, with $\Delta_{\rm vir}=(18\pi^2+82x-39x^2)/\Omega(z)$, and
$\Omega(z)=\rho_m(z)/\rho_{\rm crit}$ and $x=\Omega(z)-1$ \citep{Bryan-Norman98}. 

Having defined \Pcen, the cumulative probability that a halo of mass \mh\ hosts a
central galaxy with a stellar mass greater than \ms\ is simply
\begin{equation}
 \int^{\infty}_{M_{\rm *}}\Pcen d\ms,
  \label{Ncen}
\end{equation}
which coincides with the definition of the mean occupational number of central galaxies, \Nc. 
Finally, we are able to infer the mean number density of galaxies with stellar 
mass greater than \ms, that is, 
$n_{g,\rm cen}(>\ms)=\int^{\infty}_{0}\Nc\phih(\mh) d\mh$.

\subsubsection{Satellite galaxies}

Since satellites are expected to reside in subhalos, we will use a
similar approach to centrals, i.e.\ we will establish a link between
the properties of satellite galaxies to those of the subhalos.
However, in this case, one should take into account that (i) before becoming a satellite they
occupy a distinct halo, and (ii) the subhalo mass, \msub,
can be defined at the observation time (their present-day mass in our
case) or at the accretion time (the epoch when a distinct halo became
a subhalo).

Item (ii) is discussed in \citetalias{RDA12}.  First,
\citetalias{RDA12} show that once the subhalo mass function is
provided for any definition of subhalo mass, the satellite-to-subhalo
mass relation, \ssmr, can be constrained consistently with the
observed satellite \gsmf. Therefore, the use of one or another is
subject to practical criteria. On one hand, with the accretion-epoch
definition, the central and satellite mass relations are almost the
same, as observations (\citetalias{RDA12}) and simulations
(\citealp{Simha+2012}) show, and the obtained \ssmr\ for this case is
free of a potential dependence on host halo mass. Besides, the
accretion-time \msub\ definition is less sensitive to the specifics of
the halo finding algorithm than the observed-time definition.  On the
other hand, the \ssmr\ for the subhalo mass defined at accretion time
is actually a nominal relation, where the abundance matching is
carried out for the satellite \gsmf\ at the \emph{present epoch} but
for a subhalo mass function constructed for subhalos accreted at
\emph{different previous epochs}. The physical interpretation of this
nominal relation requires assumptions about the evolution of
galaxies. Instead, when matching present-day satellite abundances with
present-day subhalo abundances, the connection is direct and no
assumptions about evolution are necessary (see \citetalias{RDA12},
\S\S 4.1, for an extensive discussion).

Here, our constraints for the \ssmr\ refer to \msub\ defined at the
same epoch that the observational input is provided for, that is the
present time.  However, some results will be presented also for the 
accretion-time \msub.

For the subhalo abundance, given as the subhalo conditional mass
function, we use the results obtained in \citet[][]{Boylan-Kolchin+10}
based on the Millennium-II simulation.  It is worth noting that the
lowest subhalo masses we probe in this work ($\sim10^{10}-10^{11}
\msun$, depending on the \gsmf\ used) are around 3--4 orders of
magnitude above the mass resolution of this simulation.  The  present-day subhalo
mass is the mass enclosed within a truncation radius, which is
defined as the radius where the spherically-averaged density profile
starts to flatten or to increase with radius. The fitting formula for
the mean cumulative number of subhalos with present-day (observed)
mass \msub\ given a host halo mass \mh\ is:
\begin{equation}
  \Nsub=\mu_0\left(\frac{\mu}{\mu_1}\right)^a\exp\left[-\left(\frac{\mu}{\mu_{\rm cut}}\right)^b\right],
  \label{cmf_sub}
\end{equation}
where $\mu=\msub/\mh$ and  $\{\mu_0,\mu_1,\mu_{\rm cut},a,b\}=\{1.15^{(\log M_{\rm
h}-12.25)}, 0.010,0.096,-0.935,1.29\}$. Then, the number of subhalos of mass between
$\msub\pm d\msub/2$ residing in host halos of mass \mh\ (the \scmf), is simply
\begin{equation}
  \phisub d\msub=d\Nsub.
  \label{Nsub}
\end{equation}

The average cumulative number of subhalos reported in
\citet{Boylan-Kolchin+10}, Eq.~(\ref{cmf_sub}), was actually obtained
for MW-sized halos. However, as the authors discuss, the normalization
factor, $\mu_0$, has been found to vary with \mh, roughly 15\% per dex
in \mh.  For this reason we introduce the quantity $\mu_0=1.15^{(\log
  M_{\rm h}-12.25)}$ \citep[see also][]{Gao11}.

The difference in cosmology between the Millennium-II simulation and
ours leads to differences in the resulting abundances of subhalos of
roughly a few per cent in the amplitude of the subhalo mass function
(\citealp{Boylan-Kolchin+10}), and it has little effects on our
results (see \citetalias{RDA12}). In any case, we introduce a
correction to first order, taking advantage of the fact that
\citet{Tinker+2008} provides the distinct \hmf\ as a function of the
relevant cosmological parameters. First, the subhalo mass function is
calculated from Eqs.~(\ref{cmf_sub}) and (\ref{Nsub}) and the
\citet{Tinker+2008} \hmf\ defined for the Millenium cosmology. Then,
the "Millenium-cosmology" subhalo-to-halo mass function ratio is
calculated, $T(M)=\phi_{\rm sub,MII}(M)/\phi_{\rm h,MII}(M)$.  This
ratio is now used to recalculate the subhalo mass function for our
cosmology as $\phi_{\rm sub}(M)=T(M)\phih(M)$, where $\phih(M)$ is the
\citet{Tinker+2008} \hmf\ for our cosmology.  Finally, assuming the
same functional form for the subhalo conditional mass function
(Eq. \ref{Nsub}), with the same
$\mu_0,$
we obtain the new parameters for our cosmology from $\chi^2$ fitting
$\{\mu_1,\mu_{\rm cut},a,b\}=\{0.011,0.096,-0.935,1.342\}$.  These are
actually very close to what is reported in \citet{Boylan-Kolchin+10}.

Analogously to centrals, for constructing the satellite \gsmf\ we introduce 
the probability, $\Psub d\ms$, that a subhalo of mass \msub\ hosts a satellite galaxy with stellar mass between
$\ms\pm d\ms/2$.  
In general there is no reason for assuming $\Pcen=\Psub$\footnote{This assumption may actually lead to inconsistent
results, even for the accretion-time \msub\ definition, as shown in \citetalias{RDA12}.  
For \msub\ defined at the present time, tidal stripping affects the masses of the
subhalos producing this obviously a systematic offset between the galaxy-halo and satellite-subhalo
mass relations, which is sometimes incorporated as an assumed global offset in the \amt\ analyses
\citep[e.g.,][]{Vale+04,Weinberg+2008}. For \msub\ defined at the accretion epoch, the two
relations become actually close according to the extended \amt\ analysis of \citetalias{RDA12} or
to the results of cosmological simulations \citep{Simha+2012}, but there may be still offsets and 
differences in scatter because of the uncertain evolution of the satellites after accretion 
(see Fig. 2 in \citetalias{RDA12} and Figs. 3 and 7 below).}. 
We again adopt a log-normal form,
  \begin{eqnarray}
  \Psub d\ms=\frac{d\ms}{\sqrt{2\pi\sigma_s^2}\ms\ln(10)}\times& & \nonumber \\
  \exp\left[{-\frac{\log^2(\ms/\mcs)}{2\sigma_s^2}}\right],
  \label{cmfsat}
\end{eqnarray}
where $\sigma_s$ is the scatter (width) around the logarithm in base
10 of \mcs, the mean satellite-subhalo mass relation
(\ssmr). Similarly to centrals, we parametrize $\log\mcs$ using
Eq.~(\ref{msmh}).  The reason is because, as observations suggest, the
shape of the satellite \gsmf\ is also a Schechter-like function (e.g.,
\citetalias{YMB09}; \citealp{Yang2011}), which is easily reproduced
from the halo or subhalo mass function using the parametrization given
by Eq.~(\ref{msmh}).

The next step is to link satellites to subhalos. The most natural
way to do this is via their occupational numbers \citep[e.g.,][]{Yang+2009}.
Let $\cmfsat$ be the \cmf\ giving the mean number
of satellites of stellar mass $\ms\pm\ms/2$ residing in a host halo of mass \mh:
\begin{eqnarray}
  \cmfsat d\ms=& & \nonumber \\
  d\ms\int^{\infty}_{0}\Psub\phisub d\msub.
  \label{cmfsat}
\end{eqnarray}
The similarity with Eq.~(\ref{phig}) is not a coincidence, since
this is actually the \amt\ in its differential form but at the level 
of $\cmf$s. Integrating this over stellar mass gives the mean occupation of 
satellite galaxies in individual halos: 
\begin{equation}
  \Ns=\int^{\infty}_{M_{\rm *}}\cmfsat d\ms.
  \label{nsat}
\end{equation}
At this point we are in a position to
compute the satellite \gsmf:
\begin{equation}
  \phisat(\ms)d\ms=d\ms\int^{\infty}_{0} \cmfsat\phih(\mh) d\mh,
  \label{densat}
\end{equation}
and in the case that $\sigma_s$ is a constant,
\begin{equation}
  \phisat(\ms)d\ms=d\ms\int^{\infty}_{0}\Psub\densub(\msub) d\msub,
  \label{densat}
\end{equation}
which is the matching of satellite galaxies to subhalos. The mean
number density of satellite galaxies with stellar mass greater than
\ms\ is given by:
\begin{equation}
  \ngsat(>\ms)=\int^{\infty}_{0}\Ns\phih(\mh) d\mh.
\end{equation}

Finally, note that the relation between $\Pcen$ and $\Psub$ 
with the distribution $P(\ms|M)$ used in the standard
\amt\  (e.g., \citealp{ValeOstriker2008,Behroozi10})
is given by
\begin{equation}
P(\ms|M)=\frac{\densub(M)}{\phi_{\rm DM}(M)}P_{\rm sat}(\ms|M)+
  \frac{\phih(M)}{\phi_{\rm DM}(M)}P_{\rm cen}(\ms|M),
\end{equation}
where $\phi_{\rm DM}(M)=\densub(M)+\phih(M)$
and $M$ applies either to the distinct halo or subhalo masses. 
Then, the above equation relates the mass relation
commonly obtained through the standard \amt\ with those obtained
in this paper, 
\begin{eqnarray}
  \langle\log\ms(M)\rangle=\frac{\densub(M)}{\phi_{\rm DM}(M)}\langle\log M_{\rm *,s}(M)\rangle+& & \nonumber\\
  \frac{\phih(M)}{\phi_{\rm DM}(M)}\langle\log M_{\rm *,c}(M)\rangle,
  \label{amwr}
\end{eqnarray}
where $M_{\rm *,s}(M)$ and $M_{\rm *,c}(M)$ are the \ssmr\ and \shmr, respectively.
It is worth noting that the standard \amt\ is recovered if both 
$\Pcen$ and $\Psub$ are assumed to be $\delta-$functions. Then $n_{g,\rm
cen}(>\ms)+ \ngsat(>\ms)=n_{\rm sub}(>\mh)+ n_{\rm h}(>\mh)$. For a detailed discussion
see \citetalias{RDA12}.

\subsection{The two-point correlation function}
\label{2pcf}

So far, the galaxy-(sub)halo link is based on an extended \amt. However, 
having modeled the occupational numbers for central and satellite galaxies, we
can now introduce information related to the spatial clustering.
For convenience, we will write
$\langle N\rangle\equiv \NG$, $\langle N_c\rangle\equiv \Nc$  and $\langle N_s\rangle\equiv \Ns$.

As usual, the two-point correlation function is decomposed into two parts,
\begin{equation}
  1+\xigg=[1+\xiggh]+[1+\xigghh],
\end{equation}
where \xiggh\ describes pairs within the same halo (one-halo term), while
\xigghh\ describes pairs occupying  different haloes (two-halo term).

To compute the one-halo term, we need to count all galaxy pairs $\langle
N(N-1)\rangle/2$ separated by a distance $r\pm dr/2$ within individual halos
of mass \mh, following a pair distribution  $\lambda(r)dr$ weighted by the 
abundance of distinct halos, $\phih$, and normalized by the mean galaxy 
number density \ng,
\begin{eqnarray}
  1+\xiggh=\frac{1}{2\pi r^2n_g^2}\int_{0}^{\infty}
           \frac{\langle N(N-1)\rangle}{2}\lambda(r)\phih(\mh) d\mh.
\label{CF}           
\end{eqnarray}
The contribution to the mean number  of galaxy pairs from central-satellite
pairs and satellite-satellite pairs is given by
\begin{eqnarray}
  \frac{\langle N(N-1)\rangle}{2}\lambda(r)dr=\langle N_c\rangle\Nsat\lambda_{c,s}(r)dr& & \nonumber\\
  +\frac{\langle N_s(N_s-1)\rangle}{2}\lambda_{s,s}(r)dr.
  \label{pair}
\end{eqnarray}
We assume that central-satellite pairs follow a pair distribution function 
$\lambda_{c,s}(r)dr=4\pi\tilde{\rho}_{\rm NFW}(\mh,r) r^2dr$, where
$\tilde{\rho}_{\rm NFW}(\mh,r)$ is the normalized NFW halo density profile.   
The satellite-satellite pair distribution, $\lambda_{s,s}(r)dr$, is then the
normalized density profile convolved  with itself, that is,
$\lambda_{s,s}(r)dr=4\pi{\lambda}_{\rm NFW}(\mh,r) r^2dr$,
where ${\lambda}_{\rm NFW}$ is the NFW profile convolved with itself. 
An analytic expression for ${\lambda}_{\rm NFW}(\mh,r)$ is given by 
\citet{Sheth01}. Both $\tilde{\rho}_{\rm NFW}$ and ${\lambda}_{\rm NFW}$ 
depend on the halo concentration parameter, $c_{\rm NFW}$. N-body numerical 
simulations show that this parameter weakly anti-correlates with mass, 
$c_{\rm NFW}=a - b\times$log\mh, though with a large scatter.

\begin{table*}
\caption{Constraints}
\begin{center}
\begin{tabular}{c|c|clclclc|c}
{\bf Constraints}:\\
\hline
Data set & Satellite \gsmf\ & Central \gsmf\ & Total \gsmf\ & \pcf  & $\chi^2$ \\
\hline
A & YMB09 & YMB09 & ---  & \tickNo & $ \chi^2(\phi^{\rm YMB09}_{\rm sat})+\chi^2(\phi^{\rm YMB09}_{\rm cen})$\\ 
B & \tickNo  & \tickNo & YMB09 & Y11 &$\chi^2(\phi^{\rm YMB09}_{\rm all})+\chi^2(w_{\rm p,bin}^{\rm Y11})$\\ 
C & YMB09 & YMB09 & --- &  Y11  &$\chi^2(\phi^{\rm YMB09}_{\rm sat})+\chi^2(\phi^{\rm YMB09}_{\rm cen})+\chi^2(w_{\rm p,bin}^{\rm Y11})$\\   
B1 & \tickNo & \tickNo & BGD08  & Y11 &$\chi^2(\phi^{\rm BGD08}_{\rm all})+\chi^2(w_{\rm p,bin}^{\rm Y11})$\\
\hline
\hline
{\bf Predictions}:\\
\hline
Data set & Satellite \gsmf\ & Central \gsmf\ & Total \gsmf\ & \pcf  & sat. CSMF & \shmr\ \& \ssmr\ \\
\hline
A & \tickNo & \tickNo & --- & \checkmark & \checkmark & \checkmark \\ 
B & \checkmark  & \checkmark  & \tickNo & \tickNo & \checkmark & \checkmark  \\ 
C & \tickNo  & \tickNo &  \tickNo &  \tickNo &  \checkmark & \checkmark \\   
B1 & \checkmark  & \checkmark  & \tickNo & \tickNo &  \checkmark & \checkmark \\
 \hline
\end{tabular}
\end{center}
\label{constraints}
\end{table*}

Based on results of $N$-body \citep{Kravtsov04} and hydrodynamic \citep{Zheng05} 
simulations, we will assume that the number of satellite-satellite pairs follow a Poisson 
distribution with mean $\Nsat^2=\langle N_s(N_s-1)\rangle$. This is also supported by the
analysis based on a large catalog of galaxy groups by \citet{Yang+2008}.

For the two-halo term, where $r>2R_h(\mh)$, all pairs must come from galaxies
in separate halos. We compute the two-halo term from the non-linear 
matter correlation function, $\xi_m(r)$ following \citep{Smith03}:
\begin{equation}
  \xigghh=b_g^2\zeta^2(r)\xi_m(r),
\end{equation}
where $\zeta(r)$ is the scale dependence of dark matter halo bias
\citep[][see their Eq. B7]{Tinker05}, and,  
\begin{equation} 
  b_g= \frac{1}{\ng}\int_{0}^{\infty}b(\mh)\NG\phih(\mh) d\mh, 
\end{equation} 
is the galaxy bias with $b(\mh)$ being the halo bias function \citep{Sheth03}.

Once we have calculated \xigg, we relate it to the projected two-point correlation 
function (\pcf), \wp, by
\begin{equation}
  \wp=2\int_{0}^{\infty}\xi_{\rm gg}(\sqrt{r_{\rm p}^2+x^2})dx.
  \label{P-CF}
\end{equation}

In this model, Eqs.~(\ref{CF}--\ref{P-CF}) relate the observed \pcf\
to the central and satellite occupational number distributions, which
on their own are related to the central and satellite--(sub)halo mass
relations. In consequence, the correlation function is related to the
total \gsmf\ and its decomposition into centrals and satellites.
Therefore, since the \gsmf s and PCFs are tightly connected, any
combination of these 
observational constraints is not expected to provide independent
constrains on the mass relations and the occupational number
distributions.  However, we expect that the uncertainties in the
determinations of these functions are reduced as more observational
constraints are introduced. We will explore this question in more
detail in \S\S 4.1.

\subsection{Parameters in the model}

Ultimately, our model, which in total consists of ten free parameters
--if $\sigma_c$, $\sigma_s$, and the $c_{\rm NFW}-\mh$ relation are
fixed-- constrains the central and satellite stellar-to-(sub)halo mass
relations.  Five parameters are to model the \shmr\ (Eq.~\ref{msmh}):
$M_{1,c}$, $\epsilon_{c}$, $\alpha_c$, $\delta_c$, and $\gamma_c$; and
five more to model the \ssmr\ (and therefore the satellite
occupational numbers): $m_{1,s}$, $\epsilon_{s}$, $\alpha_s$,
$\delta_s$, and $\gamma_s$. Note that the success of our model relies
on the ability to choose a parametric description of the \ms--\mh\ and
\ms--\msub\ relations (Eq.~\ref{msmh}), such that the observed total
$\gsmf$ and its decomposition into centrals and satellites are
well-reproduced.  As discussed previously, the main motivation for the
functional forms chosen here is that they are able to reproduce
Schechter-like \gsmf s accurately.

Using a SDSS halo-based group catalog, \citetalias{YMB09} found that
the intrinsic scatter around the \shmr\ is approximately independent
of halo mass and log-normally distributed, with a mean width of
$\sigma_c$(log\mh)=0.173 dex.  This result is also supported by
studies of satellite-galaxy kinematics \citep{More09,More11} and
analysis using HOD models
\citep[][]{Yang+03,Cooray06,Cacciato+09,Leauthaud11B}. Additionally,
\amt\ results are able to reproduce the \gsmf\ and the spatial
clustering of galaxies simultaneously when using
$\sigma_c=$const. (\citealp[e.g.,][]{Moster10}; \citetalias{RDA12};
\citealp{Reddick+2012}).  On the other hand, the scatter $\sigma_s$
around the \ssmr\ has not yet been discussed in the literature. In
\citetalias{RDA12} it is assumed to be the same as for central
galaxies, giving results consistent with the observed projected
$\pcf$s and satellite $\cmf$s. Having said that, we assume the
intrinsic scatters $\sigma_c$ and $\sigma_s$ to be independent of halo
mass and equal to 0.173 dex. Nevertheless, as we have discussed, the
constraints provided by the \gsmf\ decomposed into centrals and
satellites and the projected \pcf s are not independent but rather
they are complementary. Therefore, when using all these constraints,
it may be possible to leave $\sigma_c$ and/or $\sigma_s$ as free
parameters.  We will perform this exercise in Section 3.3.  Finally,
for the relation of the concentration parameter $c_{\rm NFW}$ with
mass, we use the fit to numerical simulations by \citet[][]{Munoz11}.

\subsection{Observational data sets and strategy} \label{sets}

A combination of the total, central, and satellite $\gsmf$s, and the
projected $\pcf$ for different \ms\ bins are necessary to constrain
our model. In the following, we will experiment with different
combinations of these data. We wish to understand how the
stellar--(sub)halo mass relations vary depending on the combination of
observational data used to constrain them.  In particular, we would like to
explore whether the uncertainty in the model parameters drops
significantly by introducing more observational constraints.

The different observational data to be used for constraining the model 
parameters are as follows:
\begin{itemize}
\item The \citetalias{YMB09} \gsmf\ decomposed into central and
  satellite galaxies.  These data were obtained from a large
  halo-based galaxy group catalog constructed in \citet{Yang+07} from
  the SDSS DR4 (they define a central galaxy as the most massive
  galaxy in a group with the remaining galaxies being satellites).
  Both the central and the satellite \gsmf\ are well-described by
  Schechter functions, with central galaxies being the more abundant
  population at all masses, at least above the low-mass limit of the
  sample, log(\ms/\msun) = 8.4.

\item The total \citetalias{Baldry08} \gsmf, which is well described
  by a double Schechter function.  This \gsmf\ is steeper at the
  low-mass end than the \citetalias{YMB09} \gsmf\ (See also \citealp{Drory+09}).
  \citetalias{Baldry08} have actually extended the \gsmf\ to a
  lower limit, log(\ms/\msun) = 7.4, by introducing a
  surface-brightness completeness correction.
 
\item The projected \pcf s determined in five \ms\ bins by
  \citet[][hereafter \citetalias{Yang2011}]{Yang2011} based on the
  SDSS DR7.
\end{itemize}

The combinations explored to constrain the model parameters
consists of four data sets:

Set A consists of the \citetalias{YMB09} central and satellite
$\gsmf$s, and is used to constrain the model parameters of our
extended \amt; in this case, the projected $\pcf$s in various mass
bins are predicted.  Set B consists of the {\it total}
\citetalias{YMB09} \gsmf\ and the \citetalias{Yang2011} projected
$\pcf$s, and this is used to constrain our full combined model; the
\gsmf\ decomposed into centrals and satellites is a prediction.  Set
B1 is similar to set B but instead of the \citetalias{YMB09} \gsmf,
the \citet[][hereafter {BGD08}]{Baldry08} \gsmf\ is used.  Set C
consists of all the available data: the \citetalias{YMB09} \gsmf\ {\it
  decomposed into centrals/satellites} and the \citetalias{Yang2011}
projected \pcf\ determined in different \ms\ bins; this data set
over-constrains our full combined model.

We notice that fiber collisions in the SDSS data underlying the group
catalog may introduce an extra source of uncertainty when using the
satellite \gsmf\ for constraining the parameters in sets A and
C. However, this seems to be a small effect at most since
\citetalias{YMB09} show that satellite $\cmf$s with a correction for
fiber collisions are only marginally different.  It is also important
to highlight that the authors report only the diagonal elements of the
covariance matrix for the projected \pcf s. We expect that the the
full covariance matrix would reduce possible systematic errors and
extra uncertainties in some of the constrained parameters.  As
discussed in \S3.1, the lack of the covariance matrix seems to affect
the results for the abundance of satellite galaxies, however, 
these effects are of minor importance.

Table~\ref{constraints} summarizes the different data sets presented above
and specifies where the observables are used as constraints and
in where they are predicted by the model. 

We use Markov Chain Monte Carlo (MCMC) methods for sampling the best
fit parameters that maximize the likelihood function
$\mathcal{L}\propto\exp(-\chi^2/2)$.  Each MCMC chain consist of
$1.5\times10^6$ elements. See Appendix~A for details on the full
procedure.

\begin{figure}
\vspace{13.3cm}
\includegraphics{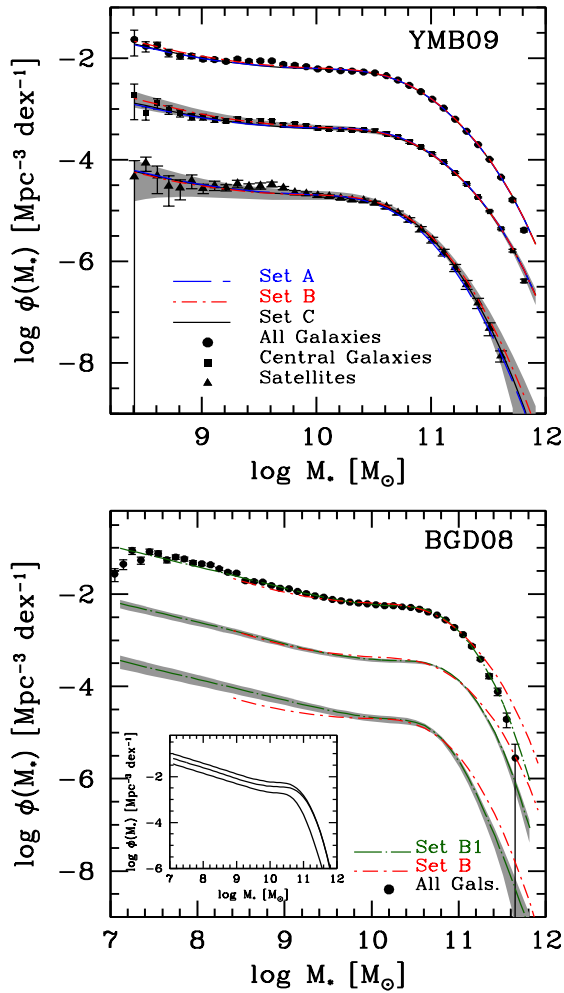}
\caption{The \gsmf\ for all, central, and satellite galaxies. For clarity,
in each panel the central \gsmf\ was shifted down by 1 dex, and the 
satellite \gsmf\ by 2 dex. \textit{Upper
panel:} Model results for sets A (blue long dashed line), 
B (red dot-short-dashed line), and C (solid line), compared
to the observed YMB09 $\gsmf$s for all (filled circles),
central (filled squares), and satellite galaxies (filled triangles).
For sets A and C the curves are just the best joint fit to the
data, while for set B are model predictions.
The shaded areas correspond to the standard deviation of the
$1.5\times10^6$ MCMC models for set B. \textit{Lower panel:} 
Same as upper panel but for the set B1 (green dot-long-dashed
line and shaded areas). The predictions for set B are 
repeated in this panel (red dot-dashed line). The corresponding
observational total \citetalias{Baldry08} \gsmf\ is showed with 
solid circles and error bars. The inset shows how the central
and the satellite $\gsmf$s add up to give the total \gsmf\ in the
case of  set B1.
 }
\label{gsmf}
\end{figure}

\begin{figure*}
\vspace*{-170pt}
\hspace*{50pt}
\includegraphics[height=6in,width=6in]{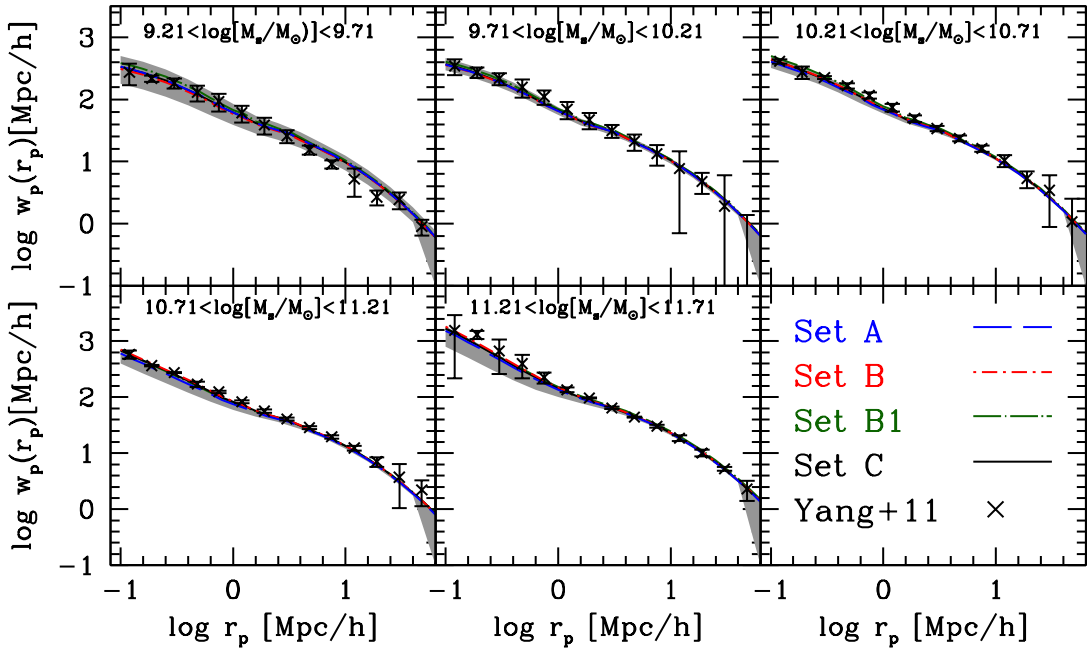}
\caption{Projected $\pcf$s in five
stellar mass bins corresponding to the data sets A, B, C, and B1 (see Table 1),
and to the \citetalias{Yang2011} observational determinations (crosses with 
error bars). The lines corresponding to each set are indicated in the last panel. 
For set A, the plotted \pcf s are predictions, while for the rest of the
sets are just the joint best fits. The shaded area show the standard 
deviation for the $1.5\times10^6$ MCMC models.}
\label{cf}
\end{figure*}

\section{The analysis}
\label{results}

In our model the central/satellite \gsmf s are tightly connected to
the projected \pcf s in such a way that given the former the latter
can be inferred and vice versa (\S\S 3.1). This connection passes
through the underlying halo/subhalo statistics and the
stellar-to-(sub)halo mass relations. Therefore, the latter, together
with the satellite \cmf, are predictions in all cases (\S\S 3.2 and \S
4, respectively).

\subsection{The $\gsmf$s \&  projected $\pcf$s}

Figure~\ref{gsmf} shows the model results for the central, satellite,
and total $\gsmf$s, while Fig.~\ref{cf} shows the projected $\pcf$s in
different mass bins.  The observational data are also plotted in these
figures (symbols with error bars).  The shaded regions in the figures
are the resulting model-fit standard deviations calculated from the
$1.5\times10^6$ MCMC models for sets B and B1 (the upper and lower
panels of Fig.~\ref{gsmf}, respectively), and for set A
(Fig.~\ref{cf}). These standard deviations are associated with the
uncertainties in the model parameters, and are produced partially by
the uncertainties in the observations used to constrain the model.
For a discussion on the model scatter see Sect.\ 5.1.

For set A, which is constrained by the \citetalias{YMB09} 
satellite and central \gsmf s (solid symbols in Fig.~\ref{gsmf}; red dot-dashed 
curves are just the joint best fits to data), the model predicts the 
projected $\pcf$s (red dot-dashed curves with shaded areas in Fig.~\ref{cf}). 
Both the amplitude and the shape of the predicted projected \pcf s are in 
excellent agreement with observations (crosses with error bars, \citetalias{Yang2011}) 
at each stellar mass bin plotted in Fig.~\ref{cf}.
This result is not surprising as shown in \citetalias{RDA12}. What is interesting, 
however, is that the standard deviations are consistent with the errors reported 
in the observations. Note that the 1-halo term is the zone with the largest 
uncertainty, which arises directly from the uncertainty in the satellite \gsmf.  
For set B, which is constrained by the total \citetalias{YMB09} \gsmf\ and the 
\citetalias{Yang2011} projected $\pcf$s (black curves are just the joint best fits 
to data), the model predicts the central and satellite $\gsmf$s 
(black curves and gray shaded regions in the upper panel of Fig.~\ref{gsmf}).  
The model predictions agree very well with observations, and therefore with set A. 
These results show that {\it the central/satellite \gsmf s and the \pcf s are tightly connected 
in such a way that given one, the other can be inferred through our model.}
Observe that in Fig.~\ref{gsmf}, the standard deviations are consistent with the error bars 
both for the satellite and the central \gsmf. The former has the largest uncertainties. 
Therefore, the lack of information from the projected \pcf\ covariance matrix seems
to affect mostly the abundance of satellite galaxies or equivalently, the 1-halo term in
the projected \pcf. 

Set B1 (lower panel of Fig.~\ref{gsmf}) is similar to set B, but the
\citetalias{Baldry08} total \gsmf\ is used as a constraint instead of
\citetalias{YMB09} data. Therefore, the model total \gsmf\ and \pcf s
(green dot-long-dashed curves in Figs.~\ref{gsmf} and \ref{cf},
respectively) are just the joint best fits to the data, but the
central and satellite \gsmf s are predicted (green dot-long-dashed
curves with gray shaded areas in the lower panel of
Fig.~\ref{gsmf}). Note that the \citetalias{Baldry08} \gsmf\ extends
to lower stellar masses. The resulting slope of the satellite \gsmf\
at the faint end ($\log(\ms/\msun)\lesssim 9.6$), is $\alpha\sim-1.5$,
which is steeper than set B, $\alpha\sim-1.2$, and the bump around
\Mstar$\sim8\times10^{10}\msun$ in the central \gsmf\ is more
pronounced.  A steeper total \gsmf\ implies a major contribution of
satellite galaxies to the total \gsmf\ at low masses.

Finally, for set C, neither the central/satellite \gsmf s nor the
projected \pcf s at different mass bins are predicted but rather employed to
constrain the model parameters. Therefore, the blue long-dashed curves
shown for set C in Figs.~\ref{gsmf} and \ref{cf} are just the joint
best fits to the observations; they are not predictions. For this set,
the predictions are the  constraints on the stellar-to-(sub)halo
mass relations. The question now is how different can these 
relations and their uncertainties be from those inferred using the
other data sets.

\subsection{Mass relations} \label{mass-rel}

\begin{table*}
\caption{Fit parameters}
\begin{center}
\begin{tabular}{|c|c|clclclclclclcl}
{\bf Central galaxies}:\\
\hline
Data set &  $\log M_{1,c}$ & stdev &$\log \epsilon_{c}$& stdev &$ \alpha_c$ & stdev &$ \delta_c$ & stdev &$ \gamma_c$ & stdev\\
\hline
A & 11.477 & 0.073 & -1.582 & 0.050 & 2.252 & 0.461 & 3.558 & 0.206 & 0.485 & 0.044 \\ 
B & 11.480 & 0.066 & -1.580 & 0.038 & 1.982 & 0.338 & 3.530 & 0.198 & 0.491 & 0.040\\ 
C & 11.493 & 0.068 & -1.600 & 0.047 & 2.138 & 0.417 & 3.572 & 0.202 & 0.487 & 0.043\\   
B1 & 11.676 & 0.056 & -1.475 & 0.027 & 2.056 & 0.110 & 2.454 & 0.183 & 0.514 & 0.047\\
\hline
\hline
{\bf Satellite galaxies}:\\
\hline
Data set &  $\log m_{1,s}$ & stdev &$ \log \epsilon_{s}$ & stdev &$ \alpha_s$ & stdev &$ \delta_s$ & stdev &$ \gamma_s$ & stdev\\
\hline
A & 10.761 & 0.069 & -0.992 & 0.063 & 2.469 & 0.710 & 3.616 & 0.260 & 0.435 & 0.077 \\ 
B & 10.773 & 0.088 & -0.951 & 0.052 & 2.670 & 0.792 & 3.612 & 0.255 & 0.437 & 0.075\\ 
C & 10.775 & 0.064 & -0.957 & 0.052 & 2.474 & 0.657 & 3.586 & 0.260 & 0.423 & 0.071\\   
B1 & 11.017 & 0.90 & -0.709 & 0.044 & 2.322 & 0.191 & 1.667 & 0.225 & 0.993 & 0.133 \\
 \hline
\end{tabular}
\end{center}
\label{parameters}
\end{table*}

\begin{figure*}
\vspace*{-220pt}
\hspace*{50pt}
\includegraphics[height=6in,width=6in]{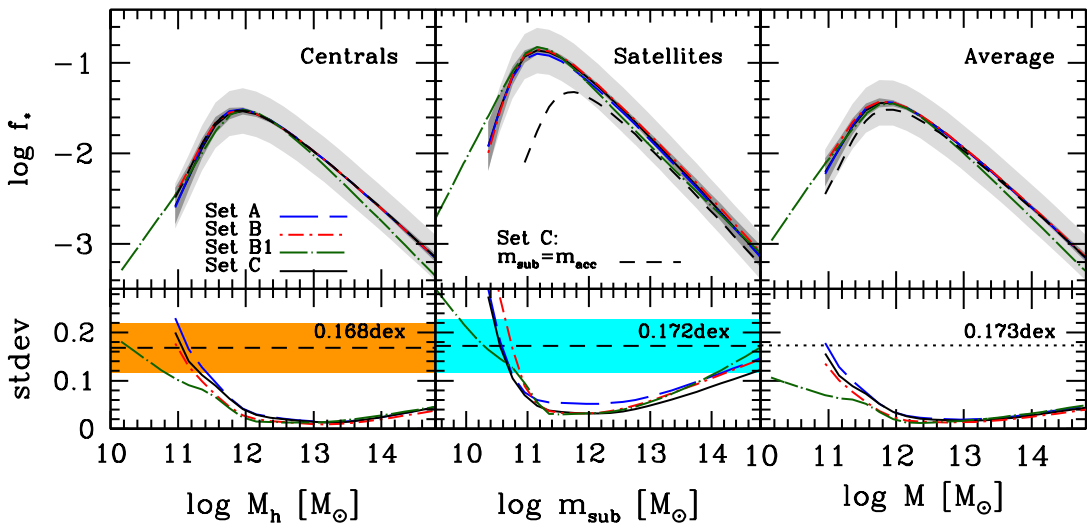}
\caption{\textit{Upper panels:} From left to right, the constrained stellar mass 
fractions of central and satellites, and of the number-density average 
(Eq. \ref{amwr}) of both. The lines corresponding to each set are
indicated inside the panels. Short-dashed curves in the second and
third panels are the constrained mass relations when the subhalo 
mass is defined at its accretion time. The systematic uncertainty
due to the uncertainty in the stellar mass determination (0.25 dex) is shown 
with the light-gray shaded areas. Gray dashed areas indicate the MCMC model-fit standard 
deviation in the case of set C. \textit{Lower panels:} The MCMC model-fit standard deviations 
for each data set. The short dashed lines in the right and middle lower panels are
the intrinsic scatters, $\sigma_c$ and $\sigma_s$, constrained for
the set C assuming them to be constant. The color shaded area show the 
standard deviations of these values.
 }
\label{shmr_app}
\end{figure*}

In the upper panels of Fig.~\ref{shmr_app} we plot the central,
satellite, and average stellar-to-(sub)halo mass ratios (stellar mass
fractions, \fs) as a function of the (sub)halo mass obtained for each
data set listed in Table~\ref{constraints}. The stellar mass fractions
are obtained directly from the corresponding mass relations. Table
\ref{parameters} lists the best fit MCMC model parameters of these
relations for each of the sets.  We use Eq.~(\ref{amwr}) to compute
the average \fs. Recall that the average relation,
$\langle\log\ms(M)\rangle$, is conceptually what is commonly obtained
with the standard \amt.  However, in the latter case it is not
possible to distinguish the mass relations for centrals and
satellites, and it is common practice to assume them equal. As
shown in \citetalias{RDA12}, this assumption is not correct.

In general, we find that the shape of the stellar fractions for both
the centrals/halos and satellites/subhalos rises steeply at low
masses, reaching a maximum and then declines roughly as a power law
towards higher masses.  We do not find significant differences among
the stellar mass fractions obtained for sets A, B and C.  Observe how
all of them lie well within the $1\sigma$ uncertainty which is
dominated by the systematic uncertainty in the stellar mass
determination ($\sim 0.25$ dex, light shaded area in
Fig.~\ref{shmr_app}; see \citealp{Behroozi10}).  All these relations
even lie well within the standard deviation of the MCMC models, shown
as a dark gray shaded area in set C (the others being very
similar). In the lower panels of Fig.~\ref{shmr_app} we plot the
the standard deviations as a function of (sub)halo mass
for each set, which we discuss in detail in \S5.1.

We arrive at two important implications: (1) the very small standard
deviations obtained for the stellar-to-(sub)halo mass relations
implies that the assumption that on average there is a monotonic
relation between galaxy and (sub)halo masses is consistent with the
data; (2) the result that set A and set B lead to very similar mass
relations confirms that {\it matching abundances is equivalent to
  matching occupational numbers and vice versa,} as suggested in
\citetalias{RDA12}.  Therefore, constraining the model parameters with
all the observational information, as in set C, should lead again to
the same stellar-to-(sub)halo mass relations as in sets A and
B. Indeed, this is what we obtain.

The central \fs--\mh\ relations for sets A--C at the low (high) mass
end scale roughly as $\fs\propto M_h^{1.5}$ ($\fs\propto
M_h^{-0.7}$). The average \fs--\mh\ relations are such that they lie
above but closer to centrals, simply because they are the dominant
population.  Instead, the satellite \fs--\msub\ relations are quite
different to centrals, both in the amplitude and in the location of
the maximum of \fs.  The maximum shifts from log(\mh/\msun)$\approx
11.9$ to log(\msub/\msun)$\approx 11.2$. These differences are
basically due to the fact that subhalos lose mass due to tidal
striping (on average 60--65\% of the mass since the accretion for
subhalos hosting satellites less massive than $\sim 2\times 10^{11}$
\msun; \citetalias{RDA12}; see also
\citealp{Vale+04,Weinberg+2008,Watson+2012b}).  
However, even when the subhalo mass at accretion time is used, some
differences remain, showing that the assumptions about evolution made
in order to construct the nominal \ssmr\ for this case are roughly
but not exactly obeyed.

In Fig.~\ref{shmr_app} we plot the model results for subhalo mass
defined at accretion time for the set C (black dashed
curve)\footnote{The \scmf\ for subhalos defined at the accretion time
  given by \citet[][see also \citealp{Giocoli+2008}]{Boylan-Kolchin+10}
  has been used (see \citetalias{RDA12} for details).}.  The
\fs--\msub\ (or \ssmr) relation now lies close to the central
\fs--\mh\ (or \shmr) relation.  Recall that for connecting the
present-day observed satellite \ms\ to (sub)halo masses 
at their different accretion epochs, one implicitly assumes that the
satellite stellar masses change in a way that at $z=0$ the \ssmr\ is
equal to the \shmr.  Our ignorance about how the satellite masses
evolve introduces an extra uncertainty in the determination of the
\ssmr\ when the subhalo mass at the accretion time is used
(\citetalias{Yang2011}; \citetalias{RDA12}). In any case, as
extensively discussed in \citetalias{RDA12}, for one or another
definition of subhalo mass, there is a unique but different average
\ssmr\ for which the satellite \gsmf\ and \cmf, and the correlation
functions are in agreement with observations.  Nevertheless, when the
\ssmr\ is assumed to be equal to the \shmr, the predicted satellite
\gsmf\ and \cmf, and correlation functions depart from
observations. They do so more strongly for the observation-time
definition, \mobs, and less strongly for the accretion-time
definition, \macc (see Figs.~1, 3, and 4 in \citetalias{RDA12}).
   
The mass relations for set B1 (green dot-dashed curves) are somewhat
different to those of set B: at high masses the central (and the
average) \fs--\mh\ relation is steeper than in set B, and at low
masses the satellite \fs--\msub\ relation is shallower.  Recall that
the massive-end of the total \citetalias{Baldry08} \gsmf\ (set B1)
decays faster than in the \citetalias{YMB09} \gsmf\ (set B; see
Fig.~\ref{gsmf}). Consequently, at a fixed \mh\ the \shmr\ for set B1
is systematically lower than for set B.  At low masses, the \gsmf\
in set B1 is steeper than in set B, causing this a steeper satellite
\gsmf\ and therefore a shallower decay of \ms\ as \msub\ decreases as
compared to set B.

\begin{figure*}
\vspace*{-170pt}
\hspace*{50pt}
\includegraphics[height=6in,width=6in]{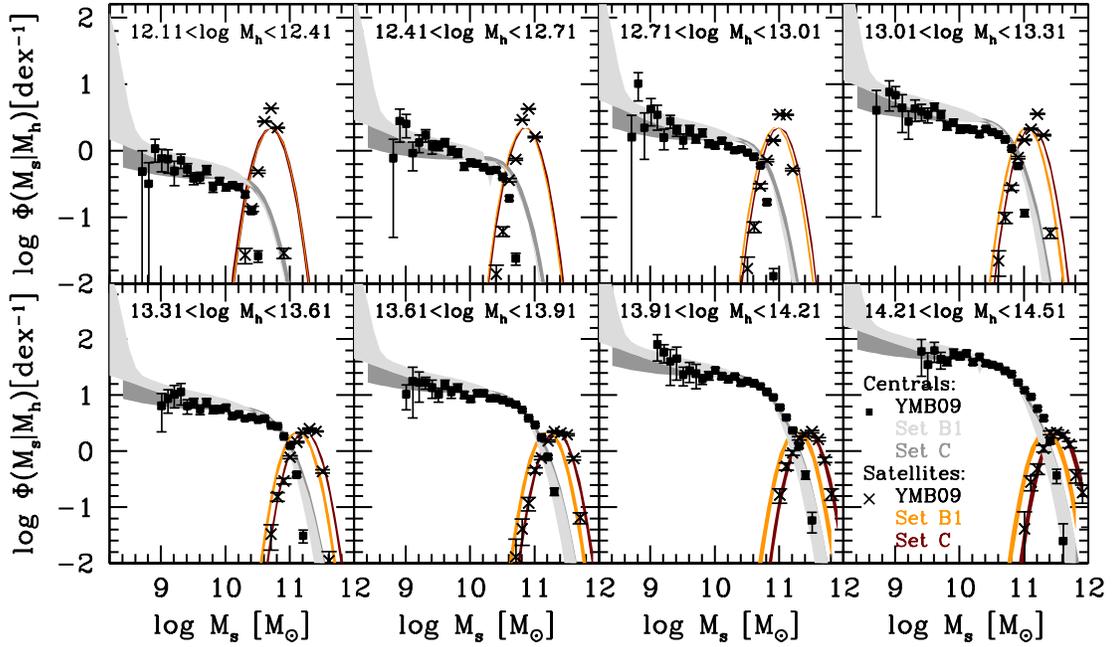}
\caption{
Central mass probability distributions and satellite $\cmf$s,  $P_{\rm cen}$ and $\Phi_{\rm sat}$, 
in eight halo mass bins for set C (dark red and dark gray shaded areas, respectively) and set B1
(orange and light gray shaded areas, respectively). The shaded areas correspond
to the standard deviation of the MCMC model-fits for
each set, which for $P_{\rm cen}$ are actually very thin. The observational 
inferences by \citetalias{YMB09} are plotted with crosses for$ P_{\rm cen}$,
and with filled squares for the satellite $\cmf$s. Their
halo masses were converted to match our virial definition.}
\label{csmf_all}
\end{figure*}

\subsection{Constraining the intrinsic scatter of the stellar-to-(sub)halo mass relations}
\label{intrinsic-scatter}

The results presented above are obtained under the assumption of
lognormal intrinsic scatter around the \shmr\ and the \ssmr\ with
constant $1\sigma$ widths of 0.173 dex for both relations.  Our model
is over-constrained by observations in set C. Therefore, we may leave
one or both of the intrinsic scatters as free parameters, but keep the
assumption that they are constant, that is, independent of (sub)halo mass.

The results of leaving only one of $\sigma_c$ or $\sigma_s$ free are
very similar to leaving both free at the same time. The MCMC algorithm
in the latter case constrains the intrinsic scatters to be $\sigma_c =
0.168\pm 0.051$ dex and $\sigma_s = 0.172\pm0.057$ dex. These values
are surprisingly close to those we have assumed.  These values are
plotted in the lower panels of Fig. \ref{shmr_app}. The constrained
mass relation parameters also remain almost the same.  So, under the
assumption of lognormal distributed and constant intrinsic scatters,
our results confirm previously estimated values of the scatter for
central galaxies, and predict similar values for the scatter around
the mass relation of satellite galaxies.

\section{Occupational statistics}

The mass relations we obtain in the previous section allow us to
explore several implications that come about naturally within the
framework of our model, in particular the halo occupational
statistics. In this section we study the implications for the
conditional stellar mass functions, the mass distribution of the most
massive satellite at a fixed halo mass, and the occurrence of
Magellanic-Clouds (MC) sized galaxies in MW-sized halos.

\subsection{The conditional stellar mass functions}
\label{csmf}

In Fig.~\ref{csmf_all}, we plot the resulting central galaxy mass probability distributions, $P_{\rm cen}$, 
and the satellite $\cmf$s, $\Phi_{\rm sat}$, in eight halo mass bins both for set C (dark red and dark gray areas, respectively) and 
set B1(orange and light gray areas, respectively). Because the predictions 
for sets A and B are very similar to those of set C we do not plot them separately. In fact, what is plotted in Fig.~\ref{csmf_all} 
are the standard deviations (scatters) of the MCMC models for
each set, which for $P_{\rm cen}$ are actually very small. 
$P_{\rm cen}$ is the probability distribution for a halo of a fixed 
mass to host a central galaxy of a given stellar mass (eq. \ref{cmfc}), while 
$\Phi_{\rm sat}$ refers to the mean number of satellite galaxies 
residing in a host halo of a fixed mass (eq. \ref{cmfsat}). 
We compute $P_{\rm cen}$ averaged in each $[M_{\rm h_1},M_{\rm h_2}]$ bin as:
\begin{equation}
  \langle P_{\rm cen}\rangle=\frac{\int_{M_{\rm h_1}}^{M_{\rm h_2}}
          \Pcen\phih(\mh) d\mh} {\int_{M_{\rm h_1}}^{M_{\rm h_2}}
          \phih(\mh) d\mh},
\end{equation}
while for satellites, the averaged \cmf\ is given by:
\begin{equation} 
  \langle \Phi_{\rm sat}\rangle=\frac{\int_{M_{\rm h_1}}^{M_{\rm h_2}} 
          \cmfsat\phih(\mh) d\mh} 
          {\int_{M_{\rm h_1}}^{M_{\rm h_2}}\phih(\mh) d\mh}. 
\end{equation} 
As seen in Fig.~\ref{csmf_all}, the smaller the halo mass, the larger is the stellar mass gap
between the most common central galaxy and the most abundant satellites.
In other words, on average, as smaller is the halo, the larger is the ratio of the central
galaxy mass to the masses of the satellite population.

In Fig.~\ref{csmf_all} we also show the corresponding observational
results by \citetalias{YMB09} for centrals (crosses) and satellites
(filled squares).  The agreement between the model predictions for set
C and the observational data is remarkable. However, some marginal
differences are observed. As the halo mass decreases, the width of the
central probability distribution is systematically somewhat broader,
and therefore its amplitude is lower compared with the
\citetalias{YMB09} data. This could be due to the assumption that the
intrinsic scatter $\sigma_c$ is independent of \mh. There are some
pieces of evidence that $\sigma_c$ slightly depends on \mh\ as
discussed in \citet[][see also \citealp{Zheng+2007}]{YMB09}.

Regarding the satellite $\cmf$s, the abundance of very massive
satellites is slightly, but systematically, overestimated for $\mh
\lesssim10^{13}\msun$. This was noted already in \citetalias{RDA12},
which suggest that a possible reason is due to the assumption that the
intrinsic scatter $\sigma_s$ is independent of \mh.  We have explored
this possibility, and found that as \mh\ decreases, the scatter
$\sigma_s$ does tend to zero in order to reproduce the
observations. Such a behavior is not expected at all.  Another
possibility, and the most likely, is that the \citetalias{YMB09}
satellite \gsmf\ is underestimated (\citealp[e.g.,][]{Skibba+2011};
\citetalias{RDA12}), although the results from set B indicate that the
obtained satellite \gsmf\ is consistent with observations.  In any
case, the excess of massive satellites in low mass halos does not
contribute significantly to the total mean density of galaxies.

For set B1, we observe that $P_{\rm cen}$ is shifted to slightly lower values of \ms\ as halo mass 
increases when comparing with observations (and set C). This is a consequence of the 
observed trends of the \ms--\mh\ relations between set B1 and C (see 
Fig.~\ref{shmr_app}), and it is ultimately related to the fact that the 
high-mass end of the \citetalias{Baldry08} \gsmf\ decreases faster than
the \citetalias{YMB09} \gsmf. On the other hand, the satellite \cmf s for set B1 
are slightly steeper at low stellar masses to those of set C. This is a consequence 
of  the \citetalias{Baldry08} \gsmf\ being steeper than that of \citetalias{YMB09}
at low masses. Note that the uncertainty in the \cmf s for set B1
dramatically increases at the lowest masses. This is because at these masses there is no
information on the \pcf, so that the total \gsmf\ alone poorly constraints the $\cmf$s. 
Set B1 also overestimates the abundance of massive satellites in low mass halos.

\subsection{Probability distributions of satellites}

Once we have constrained the distribution of satellite galaxies,
we can predict the probabilities of having $N$ satellites of a fixed
\ms\ or in a particular \ms\ range as a function of \mh. 
It is assumed in our model that the second moment of the satellite distribution
follows a Poissonian distribution (see \S\S \ref{2pcf}); the second moment
is necessary to estimate chance probabilities for any given number of satellites.
The probability of finding $N$ satellites with stellar mass above \ms\ in a host halo 
of mass \mh\ is then given by:
\begin{equation}
  P(N)=\frac{N_s^{N}e^{-N_s}}{N!},
  \label{poisson}
\end{equation}
where for convenience we redefined $N_s  = \Ns$. 

\begin{figure*}
\vspace{10.cm}
\includegraphics{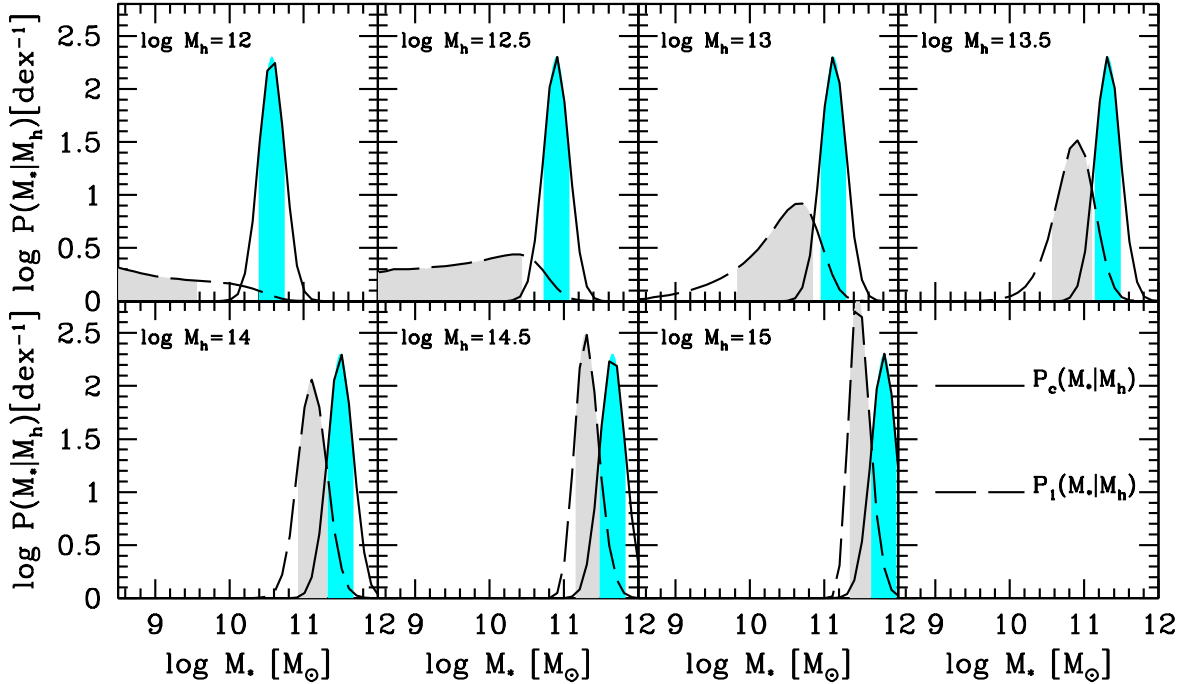}
\caption{Stellar mass distributions of centrals (solid line) and the most massive satellites
(dashed line) in seven different host halo masses. Shaded areas indicate the 68\%
of the corresponding distributions. The central galaxy masses on average are
not part of the 1$\sigma$ most-massive satellite distributions.
For masses above $\sim 3\times 10^{13}$ \msun, only approximately 15\% 
of both distributions overlap. For smaller masses, this fraction rapidly decreases
down to 3\% at $\mh=10^{12}$ \msun. The fraction of cases with overlapping 
distributions are expected to correspond to those cases where the central
galaxy mass is a statistical realization of the most-massive satellite distribution. }
\label{gap}
\end{figure*}

\subsubsection{The most massive satellite mass distribution}

We can use the satellite \cmf\ and Eq.~(\ref{poisson}) to compute the mass probability 
distribution of the most massive satellite in halos of different masses. 
This is given by the following expression \citep[e.g.,][]{Milosavljevic+2006,ValeOstriker2008}:
\begin{equation}
  \mathcal{P}_1(\ms|\mh)d\ms=\frac{\partial N_s}{\partial\ms}\times e^{-N_s}d\ms,
\label{satdistr}
\end{equation}
Note that $\int_{M_*}^{\infty}\mathcal{P}_1(\ms|\mh)d\ms=P(\geq1)$, where
$P(\geq1)$ is the probability of finding at  least one satellite galaxy more
massive than $\ms$, \mbox{$P(\geq1)=1-P(0)$}.

The results are shown in Fig.~\ref{gap}, where dashed and solid lines 
are for the most massive satellite and central galaxy mass 
distributions, respectively. The latter is by assumption a lognormal function of width
$\sigma_c= 0.173$ dex. The shaded areas indicate the $68\%$ width of the corresponding
distributions. As seen in Fig.~\ref{gap}, the mass distribution of the most massive satellite changes 
with \mh: in massive halos, it becomes closer to the distribution of the central galaxy, while in 
lower mass halos it tends towards small satellite masses compared to the central.
This difference in masses, expressed in magnitudes, is referred in the literature of galaxy 
groups/clusters as the magnitude gap. The behavior seen in Fig.~\ref{gap} is just a consequence
of the satellite \cmf s showed in Fig.~\ref{csmf_all}. 

For halos larger than $\sim1-3\times 10^{13}$ \msun, the mean and
standard deviation of the most massive satellite mass distribution
slightly increase and decrease with \mh, respectively, while for
smaller masses, the mean value of $\mathcal{P}_1(\ms|\mh)$ strongly
decreases as \mh\ decreases (faster than the central galaxy mass does)
and the standard deviation increases.  This transition is just at the
mass corresponding to small classical galaxy groups.  Therefore, our
result seems to be a consequence of the fact that in groups, the
larger the system's mass is, the smaller is the collision cross
sections for big galaxies of close masses so that more of them
survive.  Instead, in galaxy-sized halos, due to their smaller
velocity dispersions, the galaxy collision cross sections are large in
such a way that the largest galaxies probably merged into one dominant
central. Besides, the smaller the halo, the earlier most of its mass
assembled on average; hence, the (wet) mergers of the most massive
galaxies in the halo would have happened early. However, a fraction of
the galaxy-sized halos, while on average dynamically old, can accrete
massive satellites late. This could partially explain the wide
distribution of masses of the second most massive satellite in
MW-sized halos.  For example, as seen in Fig.~\ref{gap}, the
probability for these halos to have the most massive satellite $\sim
5$ times larger than the LMC is close to the probability to have this
satellite as massive as the LMC.

From Fig.~\ref{csmf_all} we see that the mass of the central galaxy in
the largest halos could be the statistical extreme of the satellite
\cmf. The question whether the brightest cluster galaxies are a mere
statistical extreme of the luminosity function in clusters or they
form a different class is a longstanding one
\citep[e.g.,][]{Tremaine+77,Hearin+2012,More2012}. According to Fig.~\ref{gap}, the mass
distributions of the most massive satellite and central galaxy become
closer as \mh\ increases. However, quantitatively, we see that the
mean of $\mathcal{P}_1(\ms|\mh)$ lies outside of the $1\sigma$ of the
central galaxy mass distribution even for a $10^{15}$ \msun\ halo,
i.e., the central and the most massive satellite galaxy, on average,
are not expected to be drawn from the same exponentially decaying mass
function; this criterion is similar to the observational one
introduced by \citet{Tremaine+77}. For a similar conclusion but using 
a different method see \citet{Hearin+2012}.

We can estimate the fraction of systems where both mass distributions
overlap, and consider that this fraction corresponds to the cases
where the most massive satellite and the central galaxy are drawn from
the same distribution.  For masses above $\sim 3\times 10^{13}$ \msun,
approximately 15\% of halos would have central galaxies that are not
statistically peculiar with respect to the satellites. For smaller
masses, this fraction rapidly decreases down to 3\% at $\mh=10^{12}$
\msun. In conclusion, most of centrals seem to form a statistically
different class of galaxies with respect to the satellites at all halo
masses, with a small fraction of cases, up to $\sim 15\%$ in
cluster-sized halos, being the exception, that is to say the centrals
in these cases could be a statistical realization of the high-mass end
of the satellite \cmf.

In order to compare our population statistics in detail with
observations, the systems should be selected by the central galaxy
\ms\ and/or group richness instead of the halo mass.  We will will
carry out this exercise elsewhere by using a mock catalog based on the
the distributions constrained with our model.

\subsubsection{The probability of Milky Way--Magellanic Clouds systems}

Our model results and Eq.~(\ref{satdistr}) can be used to compute the
probability of having one, two, or $N$ Magellanic Clouds (MCs)
satellites in MW-sized halos.  We calculate these probabilities for a
range of possible MW-halo masses discussed in the literature: $(0.7,
1, 2, 3)\times 10^{12}$ \msun.  We use $M_{\rm
  LMC}=2.3\times10^9\msun$ and $M_{\rm SMC}=5.3\times 10^8\msun$
\citep{James+11} for the stellar masses of the MCs.

\begin{figure}
\vspace*{-150pt}
\includegraphics[height=5in,width=5in]{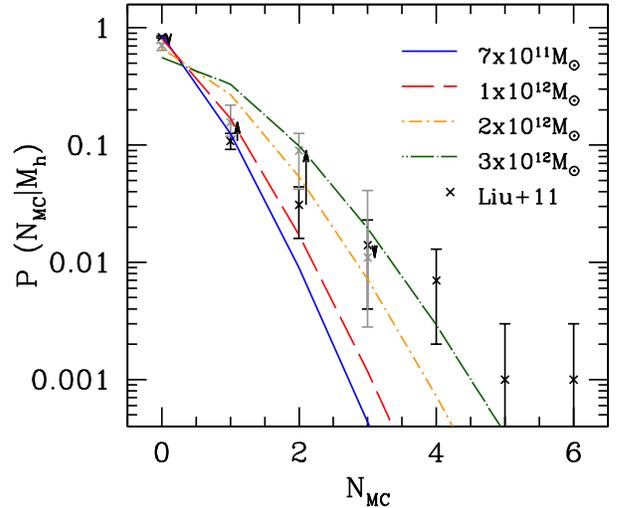}
\caption{Probability of occurrence of $N_{\rm MC}$ MC-sized
satellites in a range of possible MW-sized host halos
(different lines are for the different masses indicated in
the plot) based on the results for set C. 
Observational determinations by \citep{Liu+11} for a large sample 
of SDSS galaxies are shown with black (gray) crosses for distances 
from the host up to 150 kpc (250 kpc). 
The black arrows show how
the the occurrence of  MC-sized satellites change when the search radii 
goes from 150 kpc to 250 kpc from the host.
}
\label{pmc}
\end{figure}

Firstly, we are interested in calculating statistics that can be
compared with observations. From a large SDSS sample, \citet{Liu+11}
have estimated the fraction of isolated galaxies with MW-like
luminosities that do not have ($N_{\rm MC} = 0$) and that have $N_{\rm
  MC}=1,2,3,4,5$, or 6 MC-sized satellites.  We calculate similar
probabilities for each of the halo masses mentioned above. In order
to compare with \citet{Liu+11}, we do not exclude systems with
satellites more massive than the LMC.  The results from
\citet{Liu+11}, for a search of MC-sized satellites up to 150 kpc
around the primary, are plotted with crosses in Fig.~\ref{pmc} (from
their Table 1)\footnote{ The selection criteria and observational
  corrections for searching for MC-like satellites are actually quite
  diverse. \citet[][see also \citealp{Busha+11b}]{Liu+11} explored the
  sensitivity of the probabilities to changes in various selection
  parameters and found that their results can be slightly different,
  being the most sensitivity to the satellite search radius around the
  primary.}.  Note that in our case satellites are counted inside the
host virial radius ($\sim 200-300$ kpc).  \citet{Liu+11} plot in their
Fig.~8 the probabilities with a search radius up to 250
kpc only for $N_{\rm MC} = 0, 1, 2, 3$.  We reproduce these
measurements in Fig.~\ref{pmc} with gray symbols and error bars.
 
Our predicted probabilities for set C are plotted in Fig.~\ref{pmc}.
The probabilities of MW-like halos hosting MC-sized luminous
satellites (but including possible larger satellites) increases with
\mh. Recall that in the case of \citet{Liu+11} the central galaxy luminosity is fixed. In
this sense, our results suggest that this luminosity ($M_{\rm
  0.1r}=-21.2\pm 0.2$ mag) can be associated to halos of different
masses: for those galaxies with 1 or 2 MC-sized satellites, the
preferred masses are $\approx 1-2\times 10^{12}$ \msun, while for
those rarer systems with 3 to 6 MC-sized satellites, the preferred
masses are $>2\times 10^{12}$ \msun.  Interestingly enough, from the
inverse of the \shmr\ (set C), taking into account
the intrinsic scatter around this relation, the halo masses
corresponding to the MW stellar mass, log(\ms/\msun) = 10.74$\pm 0.1$,
are log(\mh/\msun) = $12.31\pm 0.22$. Therefore, the rare halos that
host 1 or 2 MC-sized satellites are those on the low-mass side of the
halo mass distribution given the MW central stellar mass, while the
much more rarer halos hosting 3 to 6 MC-sized satellites are those in
the high-mass end of such a distribution.

The probability of the concrete case of two MC-sized satellites ($N_{\rm MC}$=2; but not
excluding the possibility of satellites larger than the LMC) in a MW-like halo of 
$2\times 10^{12}$ \msun\ is 5.4\% for set C (see Fig.~\ref{pmc}).
If we exclude now from our model predictions the possibility of having satellites larger 
than the LMC (as it happens in the MW system), then the probability drops to a
upper limit of 2.2\%. 

The statistics of finding MW-sized galaxies with satellites in the
concrete mass range of the MCs is limited. This statistics is actually
part of the more general cumulative conditional satellite mass
function. Having this function for galaxies we may then ask, for
instance, whether the MW is rare because it has two too massive
satellites or because it has a deficiency of massive (larger than LMC)
satellites with respect to the average.  We will report results
related to these questions elsewhere by using a mock galaxy catalog
generated with the distributions constrained here. The mock catalog
will allow us also to infer several statistics given the central
galaxy stellar mass in which we are interested in (e.g., the MW one)
instead of exploring a range of possible halo masses as was done here.

We conclude that the agreement between the predicted and
observationally determined probabilities is reasonable within the
uncertainties. Such an agreement indicates that the model is
self-consist as well as consistent with the underlying \lcdm\
scenario. Note that this self-consistency has been proven down to the
scales of the MC galaxies and at the level of satellite population
distributions. Similar probabilities were found also using large
N-body cosmological simulations and looking for MW-sized halos with
subhalos that have dynamical properties similar to the MCs
\citep{Boylan-Kolchin+10, Busha+11b}.

\section{Discussion}

\subsection{Robustness and model uncertainties of the stellar-to-(sub)halo mass relations}
\label{robustness}

\begin{figure*}
\vspace*{-320pt}
\includegraphics[height=7in,width=7in]{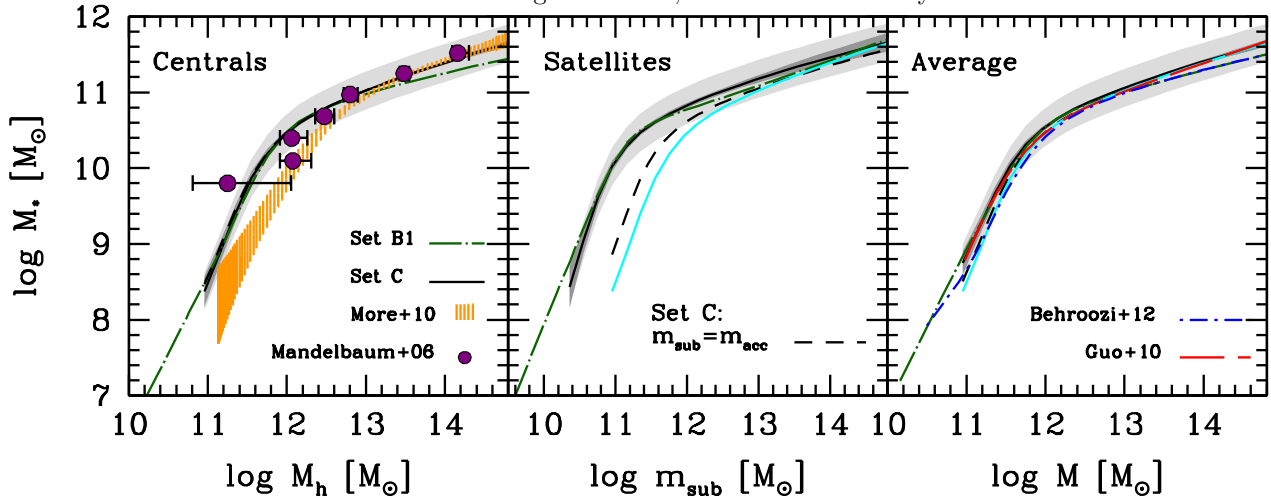}
\caption{Central-halo, satellite-subhalo, and average mass relations for
sets C (solid line) and B1 (green dot-dashed line). Short-dashed curves 
in the medium and right panels are the mass relations when
subhalo mass is defined at its accretion time. The systematic uncertainty
due to stellar mass determinations is show with the light-gray shaded area.
Gray dashed area indicates the standard deviation of the MCMC model fits in 
for set C. Filled circles with error bars correspond
to the mass relation of central galaxies from the analysis of staked weak-lensing
by \citet{Mandelbaum06}. Orange dashed area indicates the 68\% of confidence in 
the mass relation of central galaxies using the kinematics of satellites \citep{More11}. 
Abundance matching results reported in \citet{Behroozi+12},  and 
\citet{Guo10} are plotted with the blue dot-short-dashed and the red long-dashed lines. 
For comparison we have plotted with the cyan solid lines in the middle and right panels 
the central-halo mass relations for set C. Observe how in the middle panel the \ssmr\ for
the subhalo mass defined at the accretion time lies above the \shmr\ by a factor of $\sim3$, while 
in the right panel the nominal average mass relation at the accretion time is a factor of $\sim1.25$ 
higher than the  \shmr.
}
\label{shmr}
\end{figure*}

The main result from Section~\ref{results} is that both the inferred
central-halo and satellite-subhalo mass relations do not change for
all the combinations of data sets we explored.  In other words, these
relations seem to be determined robustly, no matter whether only the
central/satellite \gsmf s (set A) or whether only the total \gsmf\ and the
\pcf s (set B) or whether all of these data (set C) are used. The
results confirm what is expected: the \gsmf s of central and
satellites galaxies are well connected with the \pcf s and both are
part of a general statistical description of the galaxy
population. However, it could be that the uncertainties around the
mass relations depend on the set of observables used.  In particular,
we expect that the uncertainties should be smaller when all the
observational data are used to constrain the model.

From the results of the MCMC search over $1.5\times 10^6$ models we
can identify at each (sub)halo mass the average \ms\ and its standard
deviation.  The average stellar masses for a given (sub)halo mass are
indistinguishable to those given by the average stellar-to-(sub)halo
mass relations constructed with the best fit parameters obtained with
the MCMC method. The standard deviations can be interpreted then as
the $1\sigma$ model-fit uncertainty around these relations \citep[see
also][]{More11}.  This uncertainty is due to (i) the inability of the
proposed stellar-to-(sub)halo mass relations (Eqs.~\ref{msmh}) to
reproduce jointly the observational data, and (ii) the observational
errors in these data. The dark gray areas in Figs.~\ref{shmr_app} and
\ref{shmr} correspond to the standard deviations for the set C; the
much wider light gray areas show the scatter of 0.25 dex attributed to
the systematic uncertainty in the determination of the stellar mass
\citep{Behroozi10}.
  
How do the model-fit uncertainties in the stellar-to-(sub)halo mass
relations for set C compare with the other sets? To our surprise, the
uncertainties in sets A, B, and B1 are as small as for set
C. Actually, the uncertainties are smaller than the intrinsic scatter
between galaxy and halo mass at least for halo masses larger than
$\sim 10^{11}$ \msun.  Lower panels of Fig.~\ref{shmr_app} show the
MCMC standard deviations as a function of (sub)halo mass for all the
sets studied here.  If any, the major differences are for the
uncertainties in the \ssmr: they become large at large subhalo masses
and are larger for set A (and B1) and smaller for set C. At the
smallest (sub)halo masses the model uncertainties for all the sets
increase significantly (but yet below the systematic uncertainty of
0.25 dex). This is related to the larger observational errors at
smaller stellar masses both for the \gsmf s, in particular the one for
satellites, and the projected \pcf s.  The small model-fit
uncertainties obtained in the determination of the central-halo and
satellite-subhalo mass relations through our model again lead us to
conclude that these determinations are robust.

How do the stellar-to-(sub)halo mass relations compare with previous
work? The most direct (but highly uncertain) methods to infer the halo
masses of galaxies for large samples of objects are through weak
lensing and satellite kinematics.  In the left panel of
Fig.~\ref{shmr}, we reproduce the results for {\it central} galaxies
of $\langle\log(\ms)\rangle$ as a function of \mh\ by using stacked
satellite kinematics \citep[][Dr.\ S.\ More kindly provided us the
data in electronic form]{More11} and of $\langle\mh\rangle$ as a
function of \ms\ by using stacked weak lensing analysis\footnote{As
  widely discussed in \citet{Behroozi10}, due to the scatter, the
  inferences are slightly different depending on whether \ms\ is
  constrained as a function of \mh\ or as the inverse.  As these
  authors show the main difference is at the high-mass end.} \citep{Mandelbaum06}. The latter authors inferred the
\shmr\ separated into late- and early-type galaxies,
$\langle\mh\rangle_l$(\ms) and $\langle\mh\rangle_e$(\ms),
respectively. We compute the average $\langle\mh\rangle(\ms)$ relation
for all central galaxies as
\begin{equation}
  \langle\mh\rangle(\ms)=f_l(\ms)\langle\mh\rangle_l(\ms)+
  f_e(\ms)\langle\mh\rangle_e(\ms),
\end{equation}
where $f_l(\ms)$ and $f_e(\ms)$ are the fraction of late- and
early-type galaxies of stellar masses \ms\ in the sample.  Note also
that for the \citet{More11} and \citet{Mandelbaum06} results, small
corrections in \mh\ were applied in order to pass to our definition of
virial mass, as well as in \ms\ to be consistent with the
\citet{Chabrier2003} IMF adopted here.

As seen in Fig.~\ref{shmr}, the \citet{Mandelbaum06}, weak-lensing
determinations are consistent with our \shmr. However if one takes
into account that their dependence of $\langle\mh\rangle$ on \ms\
would be flatter at high masses in case it is deduced from the inverse
relation (see footnote), then our determination for set C would be
steeper. Instead, the results for set B1 would probably be in better
agreement with \citet{Mandelbaum06} at high masses.  The
\citet{More11} satellite-kinematics determinations are consistent with
our results for masses larger than $\mh\sim 4\times 10^{12}$ \msun,
but at smaller masses their amplitude can be 2-3 times lower. This
discrepancy between satellite kinematics and other methods has been
noted previously
(\citealp[e.g.,][]{More11,Skibba+2011,Rodriguez-Puebla11}).

In the right panels of Fig.~\ref{shmr} we compare the average
stellar-to-halo mass relation (Eq.~\ref{amwr}) obtained for set C with
those of \citet{Guo10} (red long-dashed curve) and by
\citet{Behroozi+12} (blue dot-dashed curve). These authors obtained
their relations by matching the abundances of all galaxies to
abundances of halos plus subhalos. In general, our average mass
relation is consistent with these previous global \amt\ results,
though a direct comparison might not be fair, because we do not assume
that the mass relations for centrals--halos and satellite--subhalos
are the same.  Also, the definition of subhalo mass used here (at the
time of observation) is different to the one used in the above papers,
who define it at the time of accretion.  Hence, we also plot our
\ssmr\ for the subhalo mass defined at the accretion time in
Fig.~\ref{shmr} (dashed curve in the central panel, set C) and the
corresponding average mass relation (dashed curve in the right panel).
The nominal \ssmr\ is close but not equal to the present-day \shmr. We
found that the nominal \ssmr\ lies above the \shmr\ at most by a factor of $\sim3$,
while the nominal average mass relation is a factor of $\sim1.25$ 
higher than the  \shmr, see also
\citet{Watson+2013} . To
establish the former relation, one should know how the \shmr\ (at the
accretion time, the satellite is yet a central galaxy and the subhalo
is a distinct halo) changes with time, and how the satellite mass
evolved since the accretion. Assuming that the \shmr\ is the same at
all epochs leads to the nominal \ssmr\ to be equal to the \shmr\
(\citetalias{RDA12}).  The fact that we find both relations to be
close (but not equal) implies then that the galaxy--halo connection
changes only little with time. This seems to be also the situation
in the cosmological simulations (\citealp{Simha+2012}; \citealp{deRossi+2013}).

For set B1, which uses the \citetalias{Baldry08} \gsmf, the \shmr\
changes slightly its slope at low masses, while the \ssmr\ becomes
systematically shallower than in the case of set C. This is because
the \citetalias{Baldry08} \gsmf\ becomes steeper at lower masses.
However, when the density-weighted average is calculated, the slope
change seen for the centrals is almost smeared out. For the
\citet{Behroozi+12} total (average) mass relation the slope change is
present, presumably because the contribution of the satellite-subhalo
mass relation is not taken into account. On the other hand, if we use
the subhalo mass at the accretion time instead that at the observation
time, then the smearing-out of the slope change is less evident.

\subsection{The satellite-subhalo mass relation at the low-mass end}

An interesting question is how to extend the \gsmf s and
stellar-to-(sub)halo mass relations towards low masses, since most
potential issues with the \lcdm\ scenario are happening at small
scales. As recently discussed by \citet{Boylan-Kolchin+12}, the \lcdm\
scenario can be compatible with the overall abundance of MW
satellites, but it predicts subhalos that are too massive (or too
concentrated) compared to dynamical observations of the brightest
dwarf spheroidal (dSph) satellites.  This can be visualized using the
dSph stellar mass vs.\ subhalo maximum circular velocity (or mass)
diagram, comparing the observations for the bright MW dSphs with
extrapolations of {\it total} (centrals+satellites) abundance matching results to low
masses. For a given \ms, the MW dSphs have subhalo circular velocities
(or masses) much larger (by $\sim1.5-2$ dex in mass) than the
extrapolated \amt\ results.

Our model has the advantage that it allows to constrain the \shmr\ and
the \ssmr\ separately (Fig.~\ref{shmr_app}). The extrapolation of the
latter only is what actually should be used for comparison with the MW
satellites. \citetalias{RDA12} show that if the faint-end
extrapolation of the \gsmf\ is as steep as $-1.6$
(\citetalias{Baldry08}; \citealp{Drory+09}) and is completely dominated by satellite
galaxies, then the \lcdm\ subhalo masses are consistent with the
subhalo masses of the observed MW dSphs. Here, masses are defined at
the estimated tidal radii of the dwarf satellites (see Fig.~2 in
\citetalias{RDA12} and references therein). By using our model, we are
able to decompose the \citetalias{Baldry08} \gsmf\ into satellites and
centrals (set B1; Fig.~\ref{gsmf}). The faint-end slope of the
satellite \gsmf\ (down to $\sim 2.5\times 10^7$ \msun) indeed
resembles the total mass function, \emph{but satellites do not
  dominate over centrals}. Therefore, the inferred \ssmr\ gives still
too large subhalo masses (by 0.3--0.4 dex) as compared with the tidal
masses of the MW dwarfs. We should note that in set B1, the
\citetalias{Yang2011} projected \pcf s are used, and for stellar
masses smaller than reported in \citetalias{Yang2011} ($\sim 1\times
10^9$ \msun), no projected \pcf\ constraints are applied. There are
some hints that the projected \pcf s of galaxies at small distances
(one-halo term, where satellites dominate) are steeper than those
measured in \citetalias{YMB09}, especially for the smallest galaxies
\citep{LW09}. If this is the case, then we can easily show that the
satellites become more abundant in the \gsmf\ and the \ssmr\ is
flatter at low masses, leading to a better agreement with the inferred
tidal (subhalo) masses for the MW dSph satellites.

\subsection{Interpreting the bump of the \gsmf}

Several interpretations of the shape of the total \gsmf\ have been  offered in
the literature \citep{Baldry08,Drory+09,LW09,Bolzonella+2010,Pozzetti+2010}. In this
section we will focus on interpreting the shape of the \gsmf\  using arguments
based on the occupation statistics of galaxies within halos. 
Looking at Fig.~\ref{gap}, it becomes apparent that as halo mass increases, the
likelihood of finding at least one satellite with a stellar mass {\em similar}
to that of the central galaxy increases rapidly. Also, the stellar mass range
covered by the satellite population is narrower and closer to that of the
central as halo mass increases. Assuming that these features of the satellite
population mass distribution are robust and have been in place since the
assembly of the central, it follows that the central's probability of growing by
accreting large (compared to itself) satellites was largest in high-mass
halos that today occupy the bump and high-mass end of the mass function.

\section{Conclusions}

An statistical model that combines the \amt\ with the HOD and \cmf\ formalisms
is presented. The model allows to constrain the central-halo and
satellite-subhalo mass relations (\shmr\ and \ssmr) separately, as well as the satellite
\cmf s inside the halos. The \lcdm\ halo mass function and subhalo conditional mass 
functions were used as input. From the observational point of view, the model works
with the total \gsmf\ and its decomposition into centrals and satellites, and 
the \pcf s. Therefore, the observations used to constrain the model can be different 
combinations of data: either the central/satellite \gsmf s (from \citetalias{YMB09}; set A), or the 
total \gsmf\ (from \citetalias{YMB09} or \citetalias{Baldry08}) and the \citetalias{Yang2011}
projected \pcf s (sets B and B1), or all the data, i.e., the \gsmf s of
centrals and satellites and the projected \pcf s (set C).  Our aim was to explore how sensitive
are the determinations of the mass relations and their uncertainties to the 
different data set used to constrain the model, as well as to test the overall consistency 
of the observations with the \lcdm\ halo/subhalo mass functions. Related to the latter,
we explored model predictions regarding some satellite number distributions.
The main conclusions we arrive at are:

$\bullet$ The constrained parameters of the \shmr\ and \ssmr\ are
almost identical for all sets of data, showing that these relations
(and therefore, also the satellite \cmf s) are robust with respect to
what combinations of data are used to constrain the model. To our
surprise, even the model-fit uncertainties in the constrained
stellar-to-(sub)halo mass relations are very similar for the different
combinations of data sets, including the one where all the data are
used (set C). These uncertainties are smaller than the assumed
intrinsic scatters (0.173 dex) for $\mh\grtsim 10^{11}$ \msun, and of
that order for smaller masses where the observational determinations
of the \gsmf s and projected \pcf s have larger errors.

$\bullet$ For set A, the projected \pcf s are predictions, while for
set B (and B1), the \gsmf\ decomposition into centrals and satellites
are predictions of the model. In each case, these predictions agree
very well with the observations. This shows that matching
central/satellite and (sub)halo \textit{abundances} (set A) is
equivalent to matching central/satellite and (sub)halo
\textit{occupational numbers}, in which case the \pcf s are necessary
(sets B, B1), and vice versa. In both cases, the \shmr\ and \ssmr\ are
intermediate relations. The key novelty in our model is that both
relations are constrained separately instead of being assumed equal.
Our results show also that the satellite/central \gsmf\ is tightly
connected to the spatial clustering of the population, both at the
level of the one- and two-halo terms, as well as with the satellite
mass functions inside the halos.

$\bullet$ For set C, neither the projected \pcf s nor the \gsmf\
decomposition are predictions, instead observational determinations of
these functions are used to constrain the model.  This allows us to
leave the widths of the intrinsic scatter around the \shmr\ and \ssmr\
(assumed independent of mass and log-normally distributed) as free
parameters.  We obtain $\sigma_c = 0.168\pm 0.051$ dex and $\sigma_s =
0.172\pm0.057$ dex. For centrals, our result confirms previous
estimates, and for satellites we find that the intrinsic scatter is
almost the same as for centrals.

$\bullet$ The satellite-subhalo mass relation, where subhalo masses
are defined at the observation time, is not equal to the central-halo
relation. For the former, the stellar mass scales as $\mh^{2.5}$ at
the low mass-end and as $\mh^{1.7}$ at the high-mass end (set C),
while for the latter, these scalings go as $\ms\propto \mh^{2.9}$ and
$\mh^{1.7}$, respectively.  This difference is mainly due to the fact
that subhalos lose mass (60-65\%) due to tidal striping.  When \msub\
is defined at the accretion time, the nominal \ssmr\ is actually close
to the \shmr\ but again not equal. The \ssmr\ lies above the \shmr\ at 
most by a factor of $\sim3$, while the average mass relation is a factor of $\sim1.25$ 
higher than the  \shmr, implying that the \shmr\ likely
changes little with time.  
 
$\bullet$ In set B1, we use the \citetalias{Baldry08} total \gsmf,
which extends to masses as lowe as log(\ms/\msun)=7.4.  This function
is steeper at the low-\ms\ end and decays faster at the highest masses
than the \citetalias{YMB09} \gsmf. Therefore, the \shmr\ and \ssmr\
are slightly different to those in set B. In particular, the lowest
masses show a slight flattening as compared to results of set B.  For
the satellites, if extrapolated to even lower masses, this implies
smaller subhalo masses for a given stellar mass than usually obtained
from the standard \amt. This is diminishing the potential problem of
too massive \lcdm\ subhalos for the bright MW dSphs.

Our model allows us to infer in a natural way any statistical
distribution for the central and satellite galaxy populations, as for
example the satellite \cmf\ and the mass distributions and
probabilities of particular subpopulations of satellites as a function
of halo mass.  The obtained satellite \cmf s in different halo mass
bins agree very well with those inferred from the SDSS halo-based
galaxy groups in \citetalias{YMB09}.  Moreover, we have explored in
particular two interesting statistics related to well-posed
astronomical problems, (1) the distribution of the stellar mass gap
between the central and the most-massive satellite galaxy as a
function of halo mass, and (2) the probabilities for MW-like halos to
have $N_{\rm MC}$ MC-sized satellites. Our conclusions regarding these
questions are:

$\bf (1)$ With decreasing halo mass, the mass distribution of the most
massive satellite as compared to the the distribution of the central
galaxy become more different and shifted to lower masses. This shows
this that the central is a statistically exceptional galaxy in the
halo (group). For masses larger than $\mh\sim 3\times 10^{13}$ \msun,
the differences become smaller but even in this case only $\sim 15$\%
of halos seem to have the most massive satellite statistically
indistinguishable from the central one, which implies that the latter
could be a mere statistical realization of the massive-end of the
satellite \cmf\ instead of realization of a different galaxy.

$\bf (2)$ For the range of halo masses in question for the MW, we find
that the probabilities to have $N_{\rm MC}$ MC-sized satellites are in
good agreement with the observational determinations by
\citet{Liu+11}. A MW-halo mass of $\lesssim 2\times 10^{12}$ \msun\
would agree better with the observational determinations for two
MC-sized satellites ($N_{\rm MC}=2$).  
When excluding the cases that satellites are larger than the LMC, the
probabilities become even lower: $<2.2$\% for $\mh=2\times 10^{12}$
\msun.

We conclude that the semi-empirical results we obtain here, both for
the central-halo and satellite-subhalo mass relations and their
intrinsic scatters, are quite robust and imply full consistency of the
\lcdm\ halo and subhalo populations with several statistical
distributions of the observed populations of central and satellite
galaxies down to $\ms\sim 10^9$ \msun.

\acknowledgments We thank the Referee, David Weinberg, for a
constructive report that helped to improve the paper. A. R-P
acknowledges a graduate student fellowship provided by CONACyT.
N.~D.\ and V.~A.\ acknowledge CONACyT grants 128556 and 167332.

\bibliographystyle{mn2efix.bst}
\bibliography{blib}

\appendix

\section{The fitting procedure}

From the fit to the data, we constrain the ten free parameters of
model by maximizing the likelihood function $\mathcal{L}\propto\exp(-\chi^2/2)$.
Table \ref{constraints} lists the different combination of observational
constrains used in this paper.

For each \gsmf, the $\chi^2$'s are defined as:
\begin{equation}
\chi^2(\phi^{\rm author}_{\rm tot, cen, sat})=\frac{1}{N_{\rm bin}}\sum_{i=1}^{N_{\rm bin}}
\left(\frac{\phig_{\rm ,model}^i-\phig_{\rm ,obs}^i}{\sigma_{\rm obs}^i}\right)^2,
\end{equation}
where $N_{\rm bin}$ is the number of bins in the 
total/central/satellite \gsmf\ reported for each author with an $i$th value
of $\phig_{\rm ,obs}^i$ and an error of $\sigma_{\rm obs}^i$. The $i$th value
of the  total/central/satellite \gsmf\ computed in the model 
is denoted as $\phig_{\rm ,model}^i$.

For the \citet{Yang2011} projected \pcf s, the $\chi^2$ is defined as:
\begin{equation}
\chi^2(w_{\rm p,bin}^{\rm Y11})=\frac{1}{N_{\rm s,bin}N_{\rm r,bin}}\sum_{i=1}^{N_{\rm s,bin}}
\sum_{j=1}^{N_{\rm r,bin}}
\left(\frac{\w_{\rm ,model}^{i,j}-\w_{\rm ,obs}^{i,j}}{\sigma_{\rm obs}^{i,j}}\right)^2,
\end{equation}
where $N_{\rm s,bin}$ is the number 
of stellar mass bins,  $N_{\rm r,bin}$ denotes the number of
bins in the \pcf, $\w_{\rm ,obs}^{i,j}(\w_{\rm ,model}^{i,j})$ is the amplitude of 
the observed (modeled) \pcf\ in the $j$th projected distance
bin of the $i$th stellar mass bin.

First, we find the set of parameters, $\avec=(a_1,...,a_n)$, that
minimizes $\chi^2$ using the Powell's directions set method in multi-dimensions, 
\citet{Press92}. Then, the resulting set of parameters is used as the starting point
to sample the parameter space with the MCMC method. In most of our cases $n=10$.
We also need to specify for each parameter a proposed distribution, which generates 
the candidate for sampling the parameter space.
We assume that each proposed distribution is Gaussian distributed.
The standard deviation for each parameter, $\sigma(a_i)$, is calculated from 
the covariance matrix. The covariance matrix
or error matrix of $\avec$ is defined as the inverse of the $n\times n$
matrix $\alpha=\epsilon^{-1}$, computed according to
\begin{equation}
\alpha_{kl}=\frac{1}{2}\frac{\partial^2\chi^2(\avec)}{\partial a_k \partial a_l}.
\label{error}
\end{equation}
Therefore, the 
standard deviations in the parameters correspond to the square 
roots of the terms in its diagonal, i.e., 
$\sigma(a_i)=\sqrt{\epsilon_{ii}}$. We consider these numbers as our best initial
guess for the diagonal covariance matrix of the model parameters. Then the
covariance matrix for the proposed matrix is computed according to the formula
given in \citep{Dunkley+2005}; $\epsilon_{ii}^{\rm p}=\frac{2.4}{n}^2\epsilon_{ii}$, with $n$
the number of parameters to be fitted.

Using these results, we sample a first chain with 100,000 models, from which we compute
the diagonal of the covariance matrix, $\epsilon_{ii}^{\rm c}$. 
If the ratio of each prior, $\sqrt{\epsilon_{ii}}$, to each element of the resulting
diagonal covariance matrix, $\sqrt{\epsilon_{ii}^{\rm c}}$, lies in the range 
$0.8\leq\sqrt{\epsilon_{ii}/\epsilon_{ii}^{\rm c}}\leq1.2$,
then we initialize a second chain with $1.5\times10^6$ elements for the model analysis;
else, we repeat the procedure $j-$times until the ratio of the covariances 
of the previous chain with the last one reachs the condition 
$0.8\leq\sqrt{\epsilon_{ii}^{j-1}/\epsilon_{ii}^{j}}\leq1.2$, that is to say, until there is not 
a sufficiently significant improvement in the standard deviations of the model parameters. 
The $j-$covariance 
matrix for the proposed distribution is given by $\epsilon_{ii}^{{\rm p},j}=\frac{2.4}{n}^2\epsilon_{ii}^{j}$. 
Then, we run a last chain with 
$1.5\times10^6$ elements for the model analysis. This procedure usually takes one or two
iterations. For all the chains, we find a convergence ratio in each parameter lower than 0.01
\citep{Dunkley+2005}.

\end{document}